\documentclass[journal,twocolumn,10pt]{IEEEtran}
\usepackage{latexsym}
\usepackage{amsmath,epsfig}
\usepackage{enumitem}
\usepackage{multirow}
\usepackage{subfigure}
\usepackage{multirow}
\usepackage{booktabs}
\usepackage{bigstrut}
\usepackage{array}
\usepackage{amssymb}
\usepackage{url}

\usepackage{float}
\usepackage{indentfirst}
\usepackage{times}
\usepackage{psfrag}
\usepackage{cite}
\usepackage{textcomp}
\usepackage{multirow}
\usepackage{multicol}
\usepackage{stfloats}

\newtheorem{Proof}{Proof}

\newtheorem{Analysis}{Analysis}

\newtheorem{theorem}{$\mathbf{Theorem}$}
\newtheorem{lemma}[theorem]{$\mathbf{Lemma}$}
\newtheorem{observation}[theorem]{$\mathbf{Observation}$}
\newtheorem{proposition}[theorem]{$\mathbf{Proposition}$}

\ifCLASSOPTIONcompsoc

  \usepackage[nocompress]{cite}
\else
  \usepackage{cite}
\fi

\ifCLASSINFOpdf

\else

\fi

\begin{document}

\title{Saliency-Guided Complexity Control \\for HEVC Decoding}

\author{Ren~Yang,~\IEEEmembership{Student Member,~IEEE,}
        Mai~Xu,~\IEEEmembership{Senior Member,~IEEE,} \\
        Zulin~Wang,~\IEEEmembership{Member,~IEEE,}
        Yiping Duan and~Xiaoming~Tao,~\IEEEmembership{Member,~IEEE}% <-this % stops a space
\IEEEcompsocitemizethanks{\IEEEcompsocthanksitem Ren Yang, Mai Xu, Zulin Wang are with the School of Electronic and Information Engineering, Beihang University,
China. Yiping Duan and Xiaoming Tao are with the Department of Electronic Engineering, Tsinghua University, China. This work was supported by NSFC under Grant number 61573037. Mai Xu is the corresponding author of this paper (e-mail: Maixu@buaa.edu.cn). }
%\protect\\

%E-mail: see http://www.michaelshell.org/contact.html
%\IEEEcompsocthanksitem J. Doe and J. Doe are with Anonymous University.}% <-this % stops a space
}

%{Shell \MakeLowercase{\textit{et al.}}: Bare Advanced Demo of IEEEtran.cls for IEEE Computer Society Journals}
\IEEEtitleabstractindextext{%

\begin{abstract}
The latest High Efficiency Video Coding (HEVC) standard significantly improves coding efficiency over its previous video coding standards. The expense of such improvement is enormous computational complexity, from both encoding and decoding sides. Since computational capability and power capacity  are diverse across portable devices, it is necessary to reduce decoding complexity to a target with tolerable quality loss, so called complexity control. This paper proposes a Saliency-Guided Complexity Control (SGCC) approach for HEVC decoding, which reduces the decoding complexity to the target with minimal perceptual quality loss. First, we establish the SGCC formulation to minimize perceptual quality loss at the constraint on reduced decoding complexity, which is achieved via disabling Deblocking Filter (DF) and simplifying Motion Compensation (MC) of some non-salient Coding Tree Units (CTUs). One important component in this formulation is the modelled relationship between decoding complexity reduction and DF disabling/MC simplification, which determines the control accuracy of our approach. Another component is the modelled relationship between quality loss and DF disabling/MC simplification, responsible for optimizing perceptual quality. By solving the SGCC formulation for a given target complexity, we can obtain the DF and MC settings of each CTU, and then decoding complexity can be reduced to the target. Finally, the experimental results validate the effectiveness of our SGCC approach, from the aspects of control performance, complexity-distortion performance, fluctuation of quality loss and subjective quality.
\end{abstract}

}

\maketitle

\IEEEdisplaynontitleabstractindextext

\IEEEpeerreviewmaketitle

\ifCLASSOPTIONcompsoc
\IEEEraisesectionheading{\section{Introduction}\label{sec:introduction}}
\else

\section{Introduction}
\label{sec:introduction}
\fi

\subsection{Background}
\IEEEPARstart{H}{igh} Efficiency Video Coding (HEVC) standard  \cite{sullivan2012overview} was officially approved in April 2013, significantly improving the efficiency of video coding. It is able to save around $60\%$ bit rates with similar subjective quality \cite{Tan2016}, compared with its former H.264/AVC standard. However, the cost of bit rate saving in HEVC is the huge computational complexity  \cite{bossen2012hevc}, from the aspects of both encoding and decoding. It is thus necessary to reduce encoding and decoding complexity of HEVC. The past couple of years have witnessed extensive works \cite{correa2015fast,tan2016fast,vanne2014efficient, pan2016fast,deng2014complexity,zhang2015machine, deng2015subjective} on encoding complexity reduction for HEVC. Unfortunately, there are relatively few approaches on reducing HEVC decoding complexity. Actually, decoding is far more common than encoding for existing coding standards including HEVC.  For example, according to \cite{dubey2014analysis}, the amount of videos encoded and uploaded to YouTube is only around 65 thousands every day, while there are about 100 millions videos are decoded and viewed everyday. The number of decoded videos is more than 1,000 times of encoded videos. Therefore, the study on complexity reduction is more urgent for decoding.

Moreover, different devices may be diverse in computational capability. For example, the computational capability of laptops (e.g, MacBook) is probably over twice higher than that of tablets (e.g., iPad) \cite{Everymac}. Therefore, HEVC decoding need to be adaptive to diverse computational capability. That is, it is necessary to study on reducing HEVC decoding complexity to a target, via developing complexity control approach. Unfortunately, to our best knowledge, there exists few works on complexity control for HEVC decoding. In this paper, we propose an efficient approach to achieve this goal.

\begin{figure}[t]
  \centering
  \centerline{\epsfig{figure=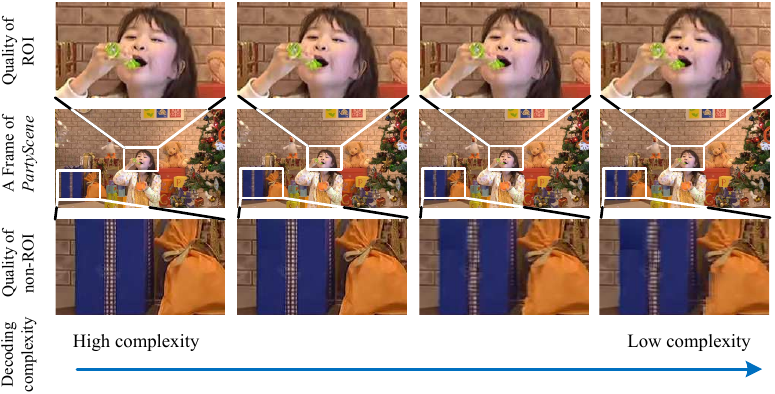, width=9cm}}
  \vspace{-0.5em}
  \caption{\footnotesize{An example of our SGCC approach. Note that each column corresponds to a specific target for HEVC decoding complexity reduction.
}}\vspace{-0.5em}
\label{motivation}
\end{figure}

\vspace{-0.5em}
\subsection{Related works}
In early time, there existed a handful of studies \cite{ma2011complexity, liu2008region} on decoding complexity reduction, for the previous H.264/AVC standard. Most recently, several approaches \cite{yan2012implementation,chi2014simd,de2015towards,chi2012parallel,alvarez2012parallel,kalali2012high,naccari2011low, feldmann2014decoder,nogues2014power,nogues2015modified} have been proposed to reduce decoding complexity/time, for the latest HEVC standard. Among them, there are two main research directions: hardware-based and algorithmic approaches.

Some works, such as \cite{yan2012implementation,chi2014simd,de2015towards,chi2012parallel,alvarez2012parallel,kalali2012high}, have been devoted to accelerating the HEVC decoding speed using hardware techniques. For example, Yan \emph{et al.} \cite{yan2012implementation} and Chi \emph{et al.} \cite{chi2014simd} proposed to take advantage of Single Instruction Multiple Data (SIMD) instructions for increasing HEVC decoding speed. Souza \emph{et al.} \cite{de2015towards} achieved the HEVC decoding acceleration, which benefits from the parallel computing of Graphics Processing Unit (GPU). Similarly, \cite{chi2012improving} presented a new parallelization approach for accelerating HEVC decoding speed with higher frame rate. The above approaches can save HEVC decoding time in some specific hardware, but they cannot reduce the complexity and power consumed by HEVC decoding. For reducing power consumption, \cite{ma2011complexity} and \cite{nogues2015dvfs} were proposed to dynamically adjust the frequency of CPU, taking advantage of Dynamic Voltage and Frequency Scaling (DVFS) technology. As such, the decoding power consumption can be reduced for H.264/AVC \cite{ma2011complexity} and HEVC \cite{nogues2015dvfs}, by means of the dynamic adjustment of CPU frequency.  In the Field-Programmable Gate Array (FPGA) platform, \cite{kalali2012high} achieved the power reduction in HEVC decoding, by designing a high-performance intra prediction hardware based on Verilog Hardware Description Language (Verilog HDL). However, all these approaches can be merely implemented on the specific hardware (e.g., GPU with SIMD, DVFS, FPGA, etc.) at the decoder side, and they are hardly adaptive to generic hardware.

For overcoming the drawback of hardware-based approaches, some algorithmic approaches have been developed to decrease video decoding complexity, via simplifying some encoding/decoding components. These approaches include \cite{liu2008region,naccari2011low,feldmann2014decoder,nogues2014power,nogues2015modified}. For H.264/AVC, Liu \emph{et al.} \cite{liu2008region} proposed to detect Region-of-Interest (ROI), and to allocate less computational resources to non-ROIs. Specifically, the total decoding complexity can be reduced with simplified coding components, according to an ROI based Rate-Distortion-Complexity (R-D-C) cost function. Later, Naccari \emph{et al.} \cite{naccari2011low} proposed an approach for reducing decoding complexity of both H.264/AVC and HEVC. In \cite{naccari2011low}, the offsets in Deblocking Filter (DF) are estimated with optimization on Generalized Block-edge Impairment Metric (GBIM), instead of  the conventional brute force optimization. This way, the computational complexity of decoding can be saved. For HEVC, the decoding complexity is reduced in \cite{feldmann2014decoder}, by modifying the structure of prediction during encoding. However, \cite{feldmann2014decoder} is not practical for already encoded videos, since it requires the modification at the encoder side. Most recently, \cite{nogues2014power} and \cite{nogues2015modified} have been proposed to modify the components at the decoder side, to make decoding complexity reduction more practical in HEVC. To be more specific, they proposed to remove some in-loop filters, and to shorten the FIR filter sizes in Motion Compensation (MC), such that HEVC decoding complexity can be reduced. In comparison with  hardware-based approaches, the algorithmic approaches on decoding complexity reduction can be implemented in any power-limited devices, but at the expense of visual quality loss.

Unfortunately, all above approaches, from both hardware-based and algorithmic aspects, cannot reduce the decoding complexity to a given target, leading to insufficient or wasteful use of power resources in some portable devices. There are only a few works on controlling decoding complexity for video coding.  For example, Langroodi \emph{et al.} \cite{langroodi2015decoder} developed a decoding complexity control approach for H.264/AVC. In \cite{langroodi2015decoder}, the decoder sends its computational resource demand to the encoder side. Then, MC is optimized at the encoder side, such that decoding complexity can be controlled at the decoder side. However, \cite{langroodi2015decoder} can be only applied to the previous H.264/AVC standard, and  it is not suitable for off-line decoding because of the communication between encoder and decode sides. To our best knowledge, there exists no approach on controlling decoding complexity for the latest HEVC standard or for off-line scenarios. More importantly, for HEVC all existing complexity reduction approaches do not take perceptual visual quality into consideration, which can be well modelled by video saliency \cite{blakemore1969existence,geisler1998real,wang2001embedded}.

\vspace{-0.5em}
\subsection{Our work and contributions}
In this paper, we propose a Saliency-Guided Complexity Control (SGCC) approach, which controls decoding complexity of HEVC, with minimization on perceptual quality loss modelled by video saliency. In our approach, we first use the method of \cite{xu2017learning} to predict video saliency map in HEVC compression domain. Then, perceptual quality is modelled, in which Mean Square Error (MSE) is weighted with the corresponding saliency values. Second, the SGCC formulation is proposed to minimize the loss of perceptual quality, when reducing HEVC decoding complexity to the target.
Since DF and MC take up large proportions in the decoding time of HEVC, the decoding complexity is reduced in our SGCC formulation by disabling DF and simplifying MC for some non-salient CTUs. Third, the relationship between decoding complexity reduction and DF disabling/MC simplification is modelled for the SGCC formulation. Similarly, the influence of DF disabling/MC simplification on visual quality is also modelled.  Finally, we develop a solution to the proposed SGCC formation, such that HEVC decoding complexity can be controlled to a target, while the perceptual quality is optimized.

Fig. \ref{motivation} shows an example of our SGCC approach. As seen in Fig. \ref{motivation}, the video quality degrades along with the reduction of decoding complexity. However, when decoding complexity reduces, our SGCC approach preserves the visual quality of ROI (e.g., face), while the quality of non-ROI degrades. As such, the perceptual quality can be optimized by our SGCC approach, when reducing decoding complexity of HEVC.

To our best knowledge, our SGCC approach is the first work to reduce decoding complexity to a target (i.e., complexity control) for HEVC, and it is also the first one to minimize perceptual quality loss in decoding complexity reduction for HEVC. This paper is an extended version of our conference paper \cite{yang2016subjective}, with extensive advanced works summarized as follows. (1) We propose to simplify MC in our new SGCC optimization formulation, with the well modelled relationship among MC simplification, quality degradation and complexity reduction. As a result, the Maximal Achievable Reduction (MAR) of HEVC decoding can increase from $\sim15\%$ to $\sim40\%$. (2) For the new SGCC optimization, an efficient solution is mathematically derived. (3) The performance of our SGCC approach is thoroughly evaluated with more test sequences, comparing approaches and evaluation metrics, than \cite{yang2016subjective}. As HEVC normally has hierarchical coding structure, temporal scalability may be applied to save some decoding complexity, which drops some upper layer frames without decoding. Our SGCC approach can be combined with temporal scalability to achieve higher reduction of decoding complexity.

\begin{table*}[!t]
\begin{center}
\footnotesize
\caption{Comparison of saliency detection performance.}\label{tab:saliency}
\begin{tabular}{|c|c|c|c|c|c|c|c|}
  \hline
         & Original \cite{xu2017learning} & Simplified \cite{xu2017learning} & PQFT~\cite{guo2010novel} & Rudoy~\cite{rudoy2013learning}& OBDL ~\cite{khatoonabadi2006many} & Itti~\cite{itti1998model}& Fang~\cite{fang2014video}\\
  \hline
  AUC &  0.80 &0.78 & 0.64 & 0.75 & 0.75 & 0.67 & 0.78\\
  \hline
  NSS & 1.27 & 1.19 & 0.48 & 1.10 & 0.96 & 0.45 & 1.26\\
  \hline
  CC & 0.42 & 0.39& 0.16 & 0.37 & 0.30 & 0.14 & 0.38\\
  \hline
\end{tabular}
\end{center}
\end{table*}
\section{Formulation for Saliency-Guided Complexity Control Approach}

\subsection{Preliminary of decoding complexity}

In \cite{bossen2012hevc}, it has been verified that DF takes up 13\%-27\% of HEVC decoding complexity (13\%-27\% for x86 and 13\%-20\% for ARM). Hence, HEVC decoding complexity can be reduced by disabling DF of some CTUs. We define $ f_n \in \{0,1\}$ to indicate whether the DF of the $n$-th CTU is enabled ($ f_n = 0$) or disabled ($ f_n = 1$). Given saliency value $w_n$ of each CTU, we define $\Delta C_D( f_n,w_n)$ as the decoding complexity reduction of a frame caused by disabling the DF of the $n$-th CTU. Note that, in this paper, the decoding complexity of HEVC is measured by the computational time on a Windows PC with Inter(R) Core(TM) i7-4790K CPU.

Also, \cite{bossen2012hevc} has investigated that MC consumes 35\%-61\% of HEVC decoding complexity (37\%-61\% for x86 and 35\%-53\% for ARM).  Thus, simplifying MC is an effective way to reduce HEVC decoding complexity. In MC, each sample of a CTU is calculated according to the corresponding samples in the reference frames. To save decoding complexity, the MC step can be skipped for some prediction samples. Instead, these samples are reconstructed by Nearest Neighbor (NN) interpolation from neighboring prediction samples, which are generated by the original MC step. In our method, for the $n$-th CTU, $g_n \in \{0,1,2,3\}$ defines that $g_n/4$ of each four samples are estimated by NN interpolation rather than applying MC. The remaining $(1-g_n/4)$ of prediction samples are decoded with the original MC step, as the reference for NN interpolation. As a result, $g_n$ implies the degree of simplifying MC. For example, $g_n = 3$ indicates the highest degree of the simplification, as $3/4$ of total samples in the $n$-th CTU skip MC. $\Delta C_{M}(g_n,w_n)$ is defined as the decoding complexity reduction of a frame, due to simplifying MC of the $n$-th CTU.

\subsection{Preliminary of saliency weighted quality}

\begin{figure}
\centering
  \includegraphics[width=.8\linewidth]{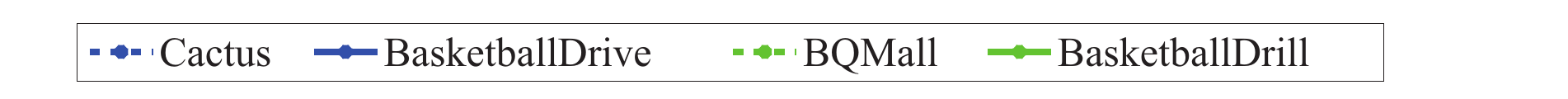}
  \vspace{.5em}
  \subfigure[Disabling DF]{\includegraphics[width=.45\linewidth]{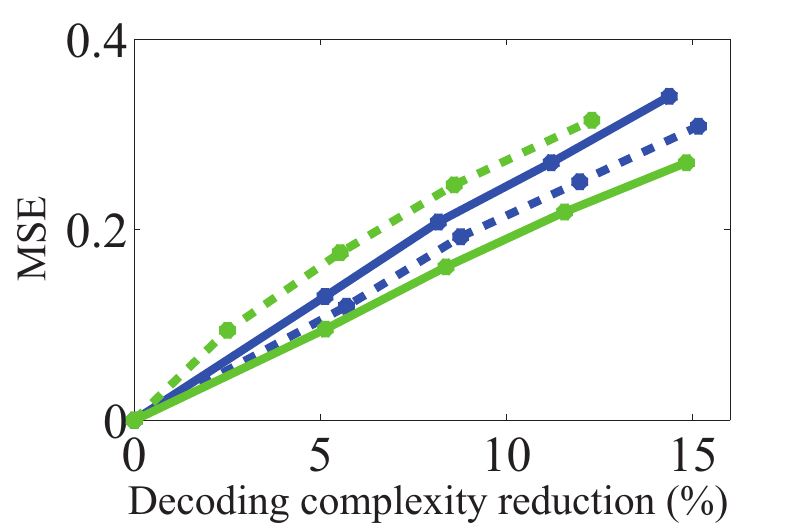}}
  \vspace{.5em}
  \subfigure[Simplifying MC]{\includegraphics[width=.45\linewidth]{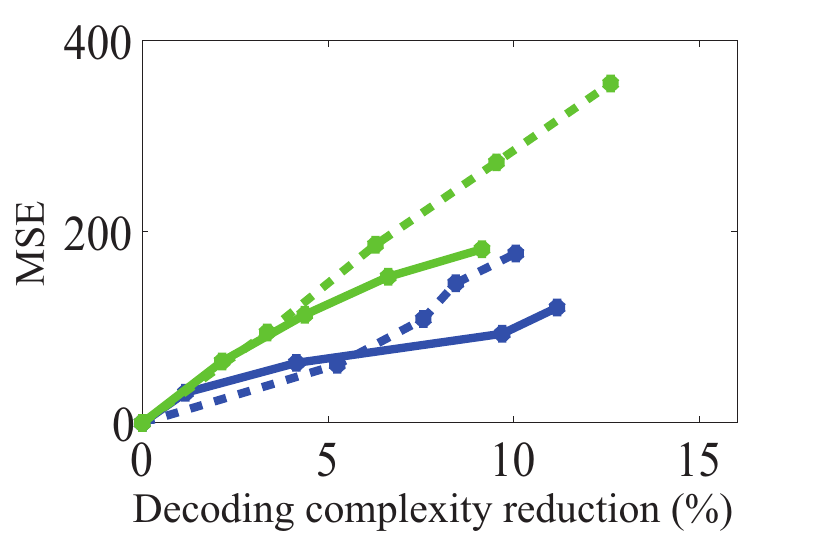}}
  \vspace{-1em}
\caption{MSE versus complexity reduction by (a) disabling DF and (b) simplifying MC for different sequences at QP = 22. Other QPs have similar results. The settings for decoding are the same as the experiments of Section \uppercase\expandafter{\romannumeral5}. The figures are obtained by setting (a) $f_n=1$ and (b) $g_n=3$ for some randomly selected CTUs, and setting $f_n=0$ and $g_n=0$ for other CTUs.}\label{fig:MSE}

\end{figure}

As the cost of decoding complexity reduction, the visual quality of decoded videos degrades (as Fig. \ref{fig:MSE} shows). Fortunately, it has been investigated \cite{geisler1998real, wang2001embedded} that visual attention of the HVS does not focus on the whole picture, but only a small region around fixation (called foveal vision). Hence, visual attention is taken into account in our approach to minimize the perceptual quality degradation.

In this paper, we use the compression domain saliency detection method \cite{xu2017learning} to directly obtain the saliency value of each CTU from HEVC bitstreams. Specifically, \cite{xu2017learning} proposed to detect saliency maps using HEVC domain features, including bits allocation, motion vectors, splitting depth and their spatial and temporal contrasts. Then, a Support Vector Machine (SVM) is learnt to combine these features for saliency detection. However, computing these complex features and applying the SVM consume high computational time, i.e., averagely 140 ms per frame. This leads to huge computational time overhead in the HEVC decoder. Since our SGCC approach aims at reducing the decoding complexity of HEVC, we simplify \cite{xu2017learning} to significantly reduce the time overhead of saliency detection.

We simplify \cite{xu2017learning} according to the two facts verified by both \cite{xu2017learning} and our experiments. (1) Bits allocation and its spatial contrast make the most essential contribution to the accuracy of saliency detection; (2) Replacing the SVM with a linear combination slightly reduces the accuracy of saliency detection, but saving extensive computational time. Therefore, the saliency value of the $n$-th CTU, defined as $w_n$, is detected by the linear combination of bits allocation and its spatial contrast, such that the computational time can be significantly reduced with slight accuracy loss of saliency detection. That is, $w_n$ is calculated by
\begin{equation}
\small
w_n = \frac{1}{2}(\frac{b_n}{b_{max}}+\frac{\Delta b_n}{\Delta b_{max}}),
\label{w}
\end{equation}
where $b_n$ and $\Delta b_{n}$ indicate allocated bits and spatial contrast of bit allocation for the $n$-th CTU, respectively. In addition, $b_{max}$ and $\Delta b_{max}$ are defined as the maximal values of $b_n$ and $\Delta b_{n}$ in a frame. In \eqref{w}, $\Delta b_{n}$ is calculated as follows,
\begin{equation}
\small
\Delta b_{n} = \left(\frac{\sum\nolimits_{n^{'}\in \textbf{I}}\exp(-\frac{d_{n'}^{2}}{2\sigma_{b}^{2}})(b_{n^{'}}-b_n)^2}{\sum\nolimits_{n^{'}\in \textbf{I}}\exp(-\frac{d_{n'}^{2}}{2\sigma_{b}^{2}})}\right)^{\frac{1}{2}},
\label{bitcontrast}
\end{equation}
where \textbf{I} is the set of 8-neighboring CTUs, and $d_{n'}$ is the Euclidean distance between the $n'$-th and $n$-th CTUs in pixel domain. After our simplification, saliency detection only consumes 0.058 ms time overhead per 1080p frame, much less than HEVC decoding complexity.

Moreover, we evaluate the performance of both the original and simplified version of \cite{xu2017learning} in terms of Area Under Receiver Operating Characteristic Curve (AUC), Normalized Scanpath Saliency (NSS) and linear Correlation Coefficient (CC) \cite{borji2013state}. The evaluation is tested over all the 15 (excluding 10-bit) sequences from Classes A-D of JCT-VC databse \cite{ohm2012comparison}. The averaged AUC, NSS and CC for the simplified version of [31] is 0.78, 1.19 and 0.39, respectively. They are slightly lower than the original version of [31] (AUC = 0.80, NSS = 1.27 and CC = 0.42), but over 2000 times computational time can be saved.

Furthermore, we also compare the performance of the simplified and original versions of \cite{xu2017learning} with other state-of-the-art saliency detection methods. The results are shown in Table \ref{tab:saliency}. It can be seen that the performance of the simplified version of \cite{xu2017learning} is comparable to or even better than other state-of-the-art saliency detection methods.

Then, we follow \cite{wang2011information}  to weight the MSE of each CTU using its saliency value. Assuming that there are in total $N$ CTUs at a video frame, the Saliency Weighted MSE (SW-MSE) of this frame is calculated by
\begin{equation}\label{sn1}
\small
  \Delta S_n( f_n,g_n,w_n) =  \frac {w_{n}}{\sum_{n=1}^{N}w_{n}}\text{MSE}(f_n,g_n).
\end{equation}
In (\ref{sn1}), $\text{MSE}(f_n,g_n)$ is the MSE between CTUs decoded by original HEVC and by HEVC with our appraoch (when the parameters are $f_n$ and $g_n$). Note that there exists $ \Delta S_n( f_n=0,g_n=0,w_n) = 0$, due to the fact that CTUs decoded by original HEVC are the same as those by our approach with $f_n=0$ and $g_n=0$. In the following, we focus on minimizing the SW-MSE when reducing decoding complexity. This way, the Quality of Experience (QoE) can be ensured.

\subsection{Formulation for SGCC approach}

Our SGCC approach aims at controlling the reduction of decoding complexity to the target, meanwhile minimizing perceptual quality loss (in terms of SW-MSE). Here, $\Delta S_n(f_n,g_n,w_n)$ and $\Delta C_n(f_n,g_n,w_n)$ are the SW-MSE and complexity reduction of the $n$-th CTU in a frame. $\Delta C_T$ is the target of complexity reduction. The optimization formulation of SGCC can be expressed by
\begin{equation}%\hspace{-.3em}
\small
  \min_{{\{ f_n,g_n\}}_{n=1}^N}\!\! \sum_{n=1}^N\!\! \Delta S_n(f_n,g_n,w_n) \ \
  \text{s.t.}\ \sum_{n=1}^N\!\! \Delta C_n(f_n,g_n,w_n)\!\! \geq\!\! \Delta C_T \label{f1},
\end{equation}
where $N$ is the total number of CTUs in a frame.

Next, we discuss how to decompose $\Delta C_n(f_n,g_n,w_n)$ and $\Delta S_n(f_n,g_n,w_n)$ in our SGCC approach, which is the first step to solve the SGCC formulation of (\ref{f1}). For the decomposition, we have the following Observations.

\begin{observation}\label{proposition1}
$\Delta C_{D}( f_n,w_n)$ and $\small \Delta C_{M}(g_n,w_n)$ are almost independent with each other. Mathematically, it holds for
\begin{equation}
\small
\sum_{n=1}^N\! \Delta C_n( f_n,g_n,w_n)\! \approx\! \sum_{n=1}^N\! \Big(\! \Delta C_{D}( f_n,w_n)\! +\! \Delta C_{M}(g_n,w_n)\!\Big). \!\!\!\!\label{c_ind}
\end{equation}
\end{observation}
\begin{Analysis}
For (\ref{c_ind}), the error rate of complexity reduction can be measured:
\begin{equation}\hspace{-0.25em}
\scriptsize
 \Delta C_e\!\! =\!\! \left|\!\frac{\sum_{n=1}^N\!\! \Delta C_n( f_n,g_n,w_n)\!\! -\!\! \sum_{n=1}^N\!\! \Big(\! \Delta C_{D}( f_n,w_n)\!\! +\!\! \Delta C_{M}(g_n,w_n)\! \Big)}{\sum_{n=1}^N\!\! \Delta C_n( f_n,g_n,w_n)}\!\right|.\!\!\!\!\!\!
\end{equation}
If $\Delta C_e\rightarrow 0$, then (\ref{c_ind}) can be obtained. Here, Table \ref{tab:cerror} reports $\Delta C_e$ of decoding several videos at QP = 22, 27, 32 and 37. Note that the settings for decoding are the same as the experiments of Section \uppercase\expandafter{\romannumeral5}. As can be seen from Table \ref{tab:cerror}, almost all average $\Delta C_e$ is less than 1.5\%. Thus, we can find $\Delta C_e\rightarrow0$, and this verifies Observation 1.  \hfill{$ \blacksquare $}
\end{Analysis}

\begin{observation}
Assume that $\Delta S_{D}(f_n,w_n)$ and $\Delta S_{M}(g_n,w_n)$ are the SW-MSEs of disabling DF and simplifying MC, respectively. They are almost independent with each other. Mathematically, it holds for
\begin{equation}
\small
\sum_{n=1}^N\! \Delta S_n( f_n,g_n,w_n)\! \approx\! \sum_{n=1}^N\! \Big(\! \Delta S_{D}( f_n,w_n)\! +\! \Delta S_{M}(g_n,w_n) \!\Big).\!\!\!\!\label{s_ind}
\end{equation}
\end{observation}
\begin{Analysis}
For (\ref{s_ind}), the error rate of SW-MSE can be measured:
\begin{equation}\hspace{-0.25em}
\scriptsize
\Delta S_e\!\! =\! \left|\!\frac {\sum_{n=1}^N\!\! \Delta\! S_n( f_n,g_n,w_n)\!\! -\!\! \sum_{n=1}^N\! \Big(\! \Delta S_{D}( f_n,w_n)\!\! +\!\! \Delta S_{M}(g_n,w_n)\! \Big)}{\sum_{n=1}^N\! \Delta\! S_n( f_n,g_n,w_n)}\!\right|.\!\!\!\!\!\!
\end{equation}

\setcounter{Proof}{2}

If $\Delta S_e\rightarrow 0$, (\ref{s_ind}) can be acquired. Table \ref{tab:serror} tabulates $\Delta S_e$ of decoding several videos at different QPs. Note that the settings are the same as the experiments of Section \uppercase\expandafter{\romannumeral5}. We can see from Table \ref{tab:serror} that most of average $\Delta S_e$ is less than 2.5\%. Thus, we can conclude that $\Delta S_e\rightarrow0$, and this verifies Observation 2.\hfill{$ \blacksquare $}
\end{Analysis}

Upon above two Observations, formulation (\ref{f1}) can be turned to
\begin{equation}\label{f2}
\small
\begin{aligned}
&\min_{{\{ f_n,g_n\}}_{n=1}^N}\sum_{n=1}^N \Big(\Delta S_D( f_n,w_n)+\Delta S_M(g_n,w_n)\Big) \\
&\text{s.t.}\quad \sum_{n=1}^N \Big(\Delta C_D( f_n,w_n)+\Delta C_M(g_n,w_n)\Big) \geq \Delta C_T.
\end{aligned}
\end{equation}
Next, we move to learn the functions of $\Delta C_D( f_n,w_n)$, $\Delta C_M(g_n,w_n)$, $\Delta S_D( f_n,w_n)$ and $\Delta S_M(g_n,w_n)$, for solving our SGCC formulation.

\begin{table}[!t]
\vspace{-2em}
\tiny
\begin{center}
\caption{Values of error rate $\Delta C_e $(\%)for all training sequences}\label{tab:cerror}
\vspace{-1em}
\begin{tabular}{|c|c|c|c|c|c|c|c|c|c|c|c|c|}
  \hline
    & \multicolumn{4}{c|}{$f_n=1, g_n=1$} &\multicolumn{4}{c|}{$f_n=1, g_n=2$} & \multicolumn{4}{c|}{$f_n=1, g_n=3$} \\
  \hline
   QP & 22 & 27 & 32 & 37& 22 & 27 & 32 & 37& 22 & 27 & 32 & 37\\
  \hline
  1 &0.7&	2.3&	2.9&	1.2&	2.6&	2.7&	0.0&	1.9&	2.0&	1.7&	0.4&	0.6\\
  \hline
  2 &0.3&	0.7&	0.4&	0.3&	3.3&	1.2&	0.3&	1.1&	1.3&	0.0&	1.0&	1.6\\
  \hline
  3 & 0.3&	0.1&	0.2&	0.9&	1.9&	1.20&	3.2&	0.4&	0.2&	2.0&	0.31&	1.5\\
  \hline
  4 & 2.1&	0.9&	1.0&	0.2&	4.8&	0.5&	0.1&	0.6&	2.3&	1.6&	2.0&	0.8\\
  \hline
  $\textbf{Ave.}$ & 0.9& 1.0&	1.1&	0.7&	3.1&	1.4&	0.5&	1.0&	1.4&	1.3&	0.9&	1.1\\
  \hline
  \multicolumn{13}{c}{\scriptsize{1: \emph{Cactus} 2: \emph{BasketballDrive} 3: \emph{BQMall} 4: \emph{BasketballDrill}}} \\
\end{tabular}
\end{center}
\vspace{-2em}
\end{table}

\section{Relationship Modelling for SGCC approach}
\subsection{Relationship modelling for $\Delta S_D(f_n,w_n)$, $\Delta S_M(g_n,w_n)$}
According to (\ref{sn1}) and Observation 2, $\Delta S_D(f_n,w_n)$ and $\Delta S_M(g_n,w_n)$ can be represented by
\begin{equation}\hspace{-2em}
\small
\begin{aligned}
\Delta S_D(f_n,w_n) &= \frac {w_{n}}{\sum_{n=1}^{N}w_{n}}\text{MSE}_D(f_n), \\
\Delta S_M(g_n,w_n) &= \frac {w_{n}}{\sum_{n=1}^{N}w_{n}}\text{MSE}_M(g_n).
\end{aligned}\label{sn22}
\end{equation}
In \eqref{sn22}, $\text{MSE}_D(f_n)$ is defined as the MSE between CTUs, decoded by our approach with $f_n\in\{0,1\}$ and by original HEVC (i.e., $f_n=0$). Similarly, $\text{MSE}_M(g_n)$ is the MSE between the CTUs decoded by our approach with $g_n\in\{0,1,2,3\}$ and by original HEVC (i.e., $g_n=0$).

It is intractable to model $\text{MSE}_D(f_n)$ and $\text{MSE}_M(g_n)$ of \eqref{sn22}, since they vary hugely across video content. However, we can use $w_n \frac {\text{MSE}_D(f_n)}{\text{MSE}_D(f_n=1)}$ and $w_{n}\frac {\text{MSE}_M(g_n)}{\text{MSE}_M(g_n=3)}$ instead of $w_n \text{MSE}_D(f_n)$ and $w_n \text{MSE}_M(g_n)$, respectively, since their correlation is rather high. Specifically, we evaluate the Spearman Rank Correlation Coefficient (SRCC) between $w_n \text{MSE}_D(f_n)$ and $ w_{n} \frac {\text{MSE}_D(f_n)}{\text{MSE}_D(f_n=1)}$ among all CTUs for each frame. The SRCC averaged over all frames of four training sequences is 0.92. Similarly, the averaged SRCC between $w_n\text{MSE}_M(g_n)$ and $w_{n}\frac {\text{MSE}_M(g_n)}{\text{MSE}_M(g_n=3)}$ is 0.70. Consequently, on the basis of \eqref{sn22}, the normalization can be written by
\begin{equation}\label{nor_sd}
\small
\Delta \widetilde S_D(f_n,w_n)\!=\!\frac {\Delta S_D(f_n,w_n)}{\Delta S_D(f_n\!=\!1,w_n\!=\!1)}\!=\!w_{n} \frac {\text{MSE}_D(f_n)}{\text{MSE}_D(f_n\!=\!1)}
\end{equation}
and
\begin{equation}\label{nor_sm}
\small
\Delta \widetilde S_M(g_n,w_n)\!=\!\frac {\Delta S_M(g_n,w_n)}{\Delta S_M(g_n\!=\!3,w_n\!=\!1)}\!=\!w_{n} \frac {\text{MSE}_M(g_n)}{\text{MSE}_M(g_n\!=\!3)},
\end{equation}
since $w_n=1$, $f_n=1$ and $g_n=3$ make SW-MSE largest in HEVC decoding. Recall that $w_n$ of each CTU can be obtained using the saliency detection method of \cite{xu2017learning}. Thus, we focus on estimating $\frac{\text{MSE}_D(f_n)}{\text{MSE}_D(f_n=1)}$ and $\frac{\text{MSE}_M(g_n)}{\text{MSE}_M(g_n=3)}$ for $\Delta \widetilde S_D(f_n,w_n)$ and $\Delta \widetilde S_M(g_n,w_n)$.

\begin{table}[!t]
\vspace{-2em}
\tiny
\begin{center}
\caption{Values of error rate $\Delta S_e $(\%) for all training sequences}\label{tab:serror}
\vspace{-1em}
\begin{tabular}{|c|c|c|c|c|c|c|c|c|c|c|c|c|}
  \hline
    & \multicolumn{4}{c|}{$f_n=1, g_n=1$} &\multicolumn{4}{c|}{$f_n=1, g_n=2$} & \multicolumn{4}{c|}{$f_n=1, g_n=3$} \\
  \hline
   QP & 22 & 27 & 32 & 37& 22 & 27 & 32 & 37& 22 & 27 & 32 & 37\\
  \hline
  1 & 1.6&2.6 &	3.5 &5.1 &1.0 &1.3 &1.6 &2.6 &0.0 &0.4&	0.7&1.0	 \\
  \hline
  2 &2.1&	2.8&	3.8&	3.1& 0.7&	1.2&	1.6&	2.1&	0.1&	0.4&	0.8&	1.2\\
  \hline
  3 &0.6  &	1.7 &2.5	&4.3 &0.3 &0.8 &1.1 &1.9 &0.1&0.4&0.7 &	1.3 \\
  \hline
  4 & 0.8&	1.6&	2.2&	3.1&	0.4&	1.2&	1.6&	2.3&	0.1&	0.5&	0.7&	1.0\\
  \hline
  $\textbf{Ave.}$ & 1.3&2.1 &3.0 &3.9&0.6 &1.1 &1.5&	2.2&0.1 &0.4&0.7 &1.1\\
  \hline
  \multicolumn{13}{c}{\scriptsize{1: \emph{Cactus} 2: \emph{BasketballDrive} 3: \emph{BQMall} 4: \emph{BasketballDrill} }} \\
\end{tabular}
\end{center}
\vspace{-2em}
\end{table}

First, we deal with the estimation on $\frac{\text{MSE}_D(f_n)}{\text{MSE}_D(f_n=1)}$. Obviously, if $f_n=1$, we have $\frac{\text{MSE}_D(f_n)}{\text{MSE}_D(f_n=1)}=1$. If $f_n=0$, DF is enabled such that we have $\frac{\text{MSE}_D(f_n)}{\text{MSE}_D(f_n=1)}=0$. Therefore, the following function holds,
\begin{equation}
\small
\frac {\text{MSE}_D(f_n)}{\text{MSE}_D(f_n=1)}=\left\{
\begin{aligned}
0&,& \quad \text{if}\ f_n=0, \\
1&,& \quad \text{if}\ f_n=1. \\
\end{aligned}
\right.\label{msefn_nor}
\end{equation}
Based on (\ref{nor_sd}) and (\ref{msefn_nor}), we can obtain
\begin{equation}\label{sd_nor}
\small
\Delta \widetilde S_D(f_n,w_n)=w_n\cdot f_n.
\end{equation}

Second, we discuss on learning $\frac{\text{MSE}_M(g_n)}{\text{MSE}_M(g_n=3)}$ from some training sequences. Four sequences, selected from JCT-VC database \cite{ohm2012comparison}, are used for training, including two  $1920\times 1080$ sequences \emph{Cactus} and \emph{BasketballDrive} from Class B, as well as two $832 \times 480$ sequences \emph{BQMall} and \emph{BasketballDrill} from Class C. The sequences are compressed by HM 16.0 at four different QPs, i.e., QP = 22, 27, 32 and 37. All settings are the same as those in Section \uppercase\expandafter{\romannumeral5}.

Four training sequences (at QP = 22, 27, 32 and 37) are decoded with MC skipped for $0$, $1$, $2$ and $3$ samples among each four samples, corresponding to $g_n=0,1,2,3$. As such, $0$, $1/4$, $1/2$ and $3/4$ of total samples are skipped for MC in each training CTU. Accordingly, for a training sequence, the MSE caused by skipping MC can be estimated by
\begin{equation}\label{mse*}
\small
\text{MSE}_M^{*}(g_n)=\frac{1}{L}\sum_{l=1}^{L}\frac{\Vert \text{I}_l(g_l=g_n)-\text{I}_l(g_l=0)\Vert_2^2}{P_l},
\end{equation}
where $\textbf{I}_l$ denotes the sample set of the $l$-th training CTU, and $L$ is the total CTU number in the training sequence. $P_l$ is the number of samples in the $l$-th training CTU, and $g_l$ denotes the proportion of its samples with MC skipped.
Given (\ref{mse*}), $\{ \text{MSE}_M^{*}(g_n)\}_{g_n=0}^3$ can be obtained for each training sequence at one QP. Afterwards, $\text{MSE}_M^*(g_n)$ is normalized by $\frac{\text{MSE}_M^*(g_n)}{\text{MSE}_M^*(g_n=3)}$. Based on the samples of $\frac{\text{MSE}_M^*(g_n)}{\text{MSE}_M^*(g_n=3)}$ for all training sequences at four QPs, we utilize the least-square fitting of the third-order polynomial regression to learn $\frac{\text{MSE}_M(g_n)}{\text{MSE}_M(g_n=3)}$.

\begin{figure}
\vspace{-1em}
  \centering
  \centerline{\epsfig{figure=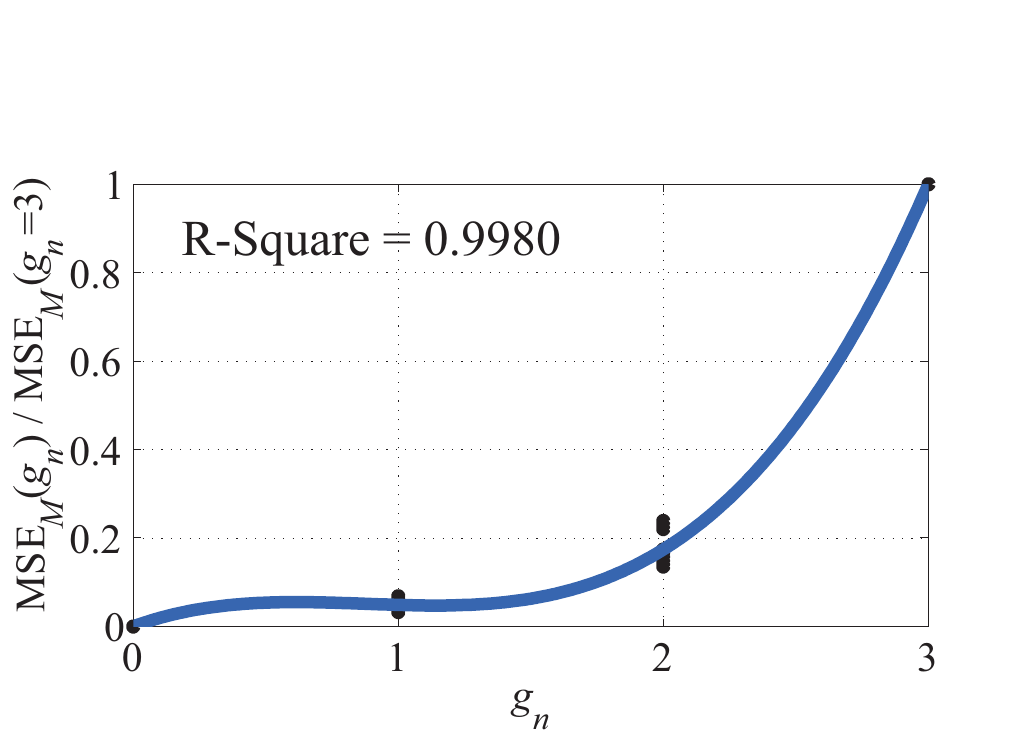, width=5cm}}
  \vspace{-1em}
  \caption{\footnotesize{Fitting curve for modelling $\frac{\text{MSE}_M(g_n)}{\text{MSE}_M(g_n=3)}$}.}
\vspace{-1.5em}
\label{fig:msefits}
\end{figure}

The fitting curve is shown in Fig. \ref{fig:msefits}, each dot of which indicates a pair of $(g_n,\frac{\text{MSE}_M^*(g_n)}{\text{MSE}_M^*(g_n=3)})$ for a training sequence at one QP. Obviously, $\frac{\text{MSE}_M^*(g_n)}{\text{MSE}_M^*(g_n=3)}=1$ for $g_n=3$ and $\frac{\text{MSE}_M^*(g_n)}{\text{MSE}_M^*(g_n=3)}=0$ for $g_n=0$, due to $\text{MSE}_M^*(g_n=0)=0$. The R-square value of the fitting in Fig. \ref{fig:msefits} is 0.9980, verifying the effectiveness of the fitting model. Finally, the learnt polynomial function is as follows,
\begin{equation}\label{1}
\small
\frac{\text{MSE}_M(g_n)}{\text{MSE}_M(g_n=3)}=h_1\cdot g_n^3+h_2\cdot g_n^2+h_3\cdot g_n,
\end{equation}
where the values of $h_1$, $h_2$ and $h_3$ are presented in Table \ref{tab:coc}.
Consequently, \eqref{nor_sm} can be turned to
\begin{equation}\label{sm_nor}\hspace{-1em}
\small
\Delta \widetilde S_M(g_n,w_n)=w_n\cdot(h_1\cdot g_n^3+h_2\cdot g_n^2+h_3\cdot g_n).
\end{equation}

\vspace{-1em}
\subsection{Relationship Modelling for $\Delta C_D(f_n,w_n)$}\label{sec:relafn}
Now, we move to the modelling of $\Delta C_D(f_n,w_n)$. Obviously, we have $\Delta C_D(f_n=0,w_n)=0$, as the decoding complexity is not reduced when DF is enabled ($f_n=0$) for the $n$-th CTU. Next, we provide a way to learn $\Delta C_D(f_n=1,w_n)$.

For learning $\Delta C_D(f_n=1,w_n)$, four training sequences at four QPs are decoded with DF enabled and disabled, respectively. Then, for the $l$-th training CTU, the training sample $\Delta C^*_D(f_l=1,w_l)$ can be calculated as the percentage of complexity reduction of a frame, after disabling the DF of the $l$-th training CTU. Here, $w_l$ is the saliency values of the $l$-th training CTU.

\begin{figure}\vspace{-1em}
\begin{center}
  \subfigure[QP = 22]{\includegraphics[width=.48\linewidth]{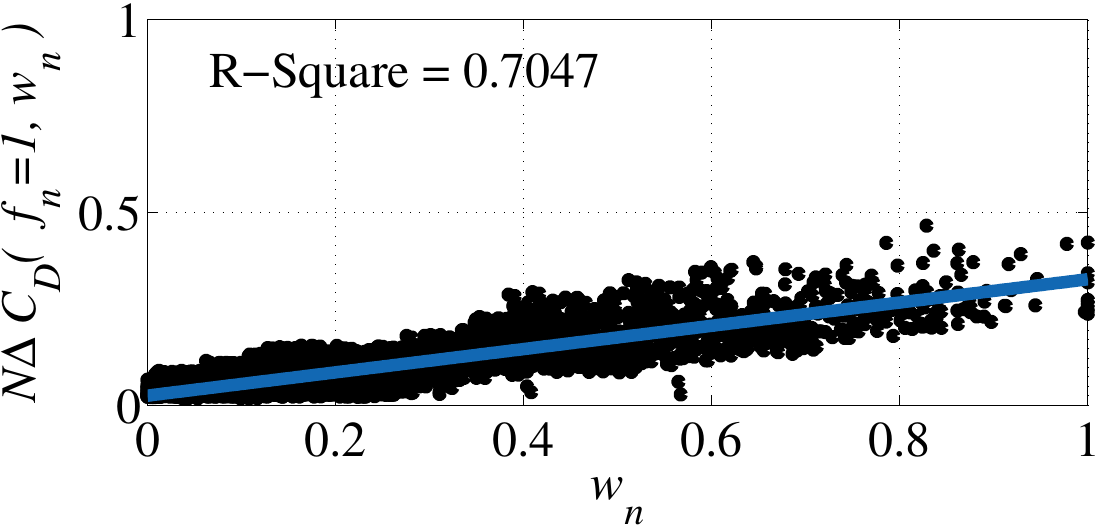}}
  \subfigure[QP = 27]{\includegraphics[width=.48\linewidth]{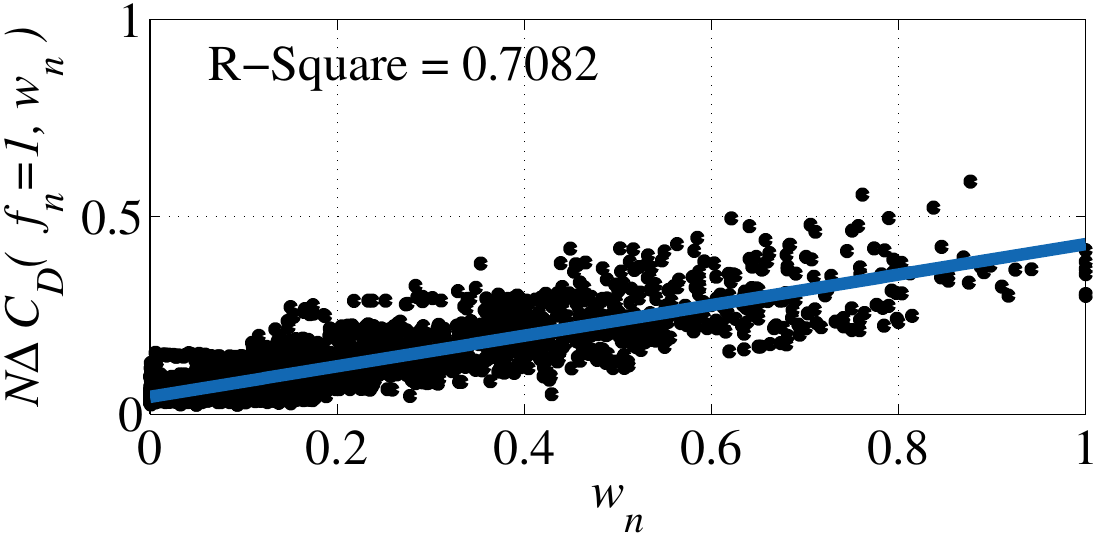}}
  \subfigure[QP = 32]{\includegraphics[width=.48\linewidth]{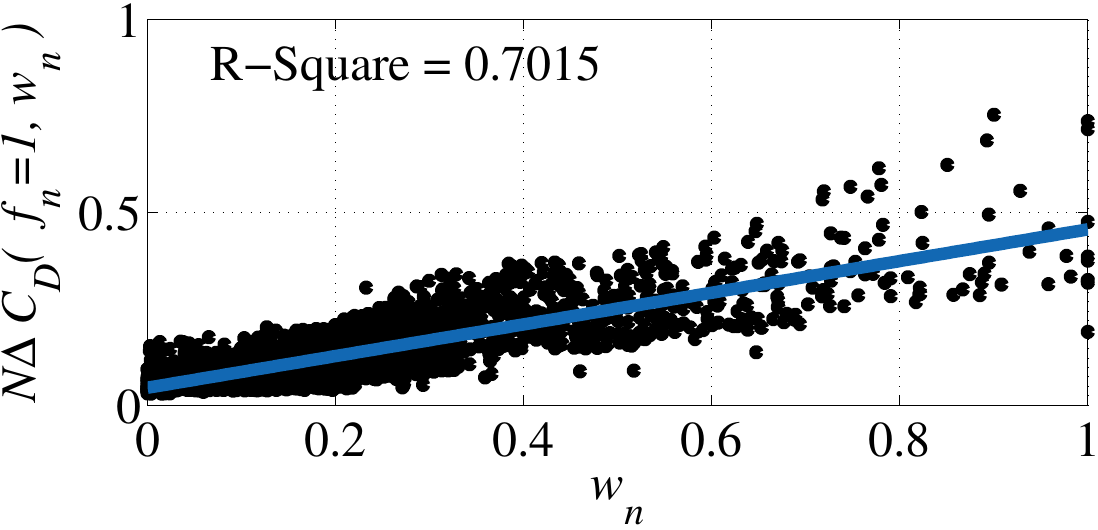}}
  \subfigure[QP = 37]{\includegraphics[width=.48\linewidth]{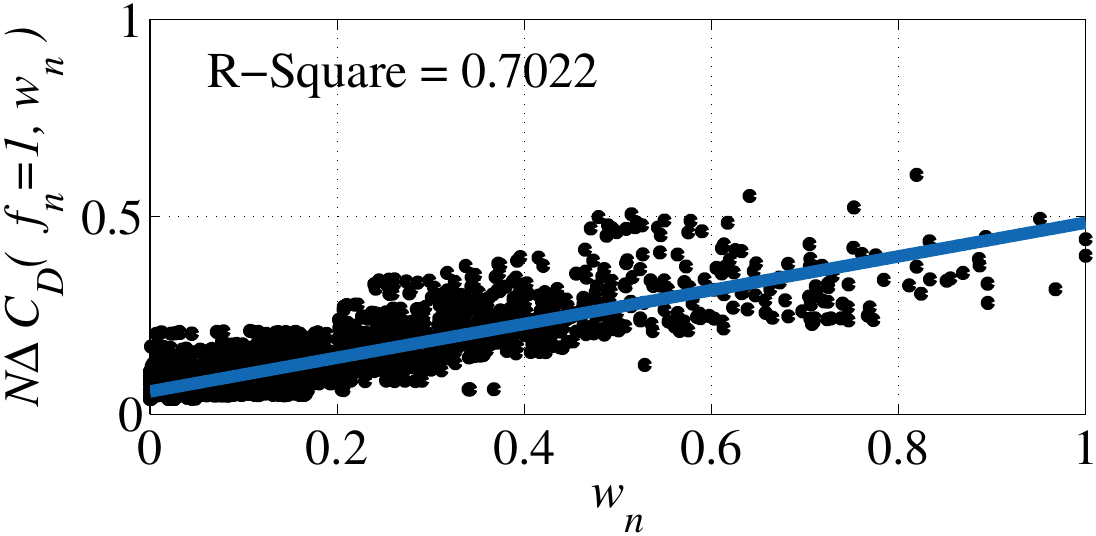}}
\vspace{-.5em}
\caption{\footnotesize{Fitting curves of $w_n$ versus $N\Delta C_D(f_n=1,w_n)$. }}\label{fig:fits}
\end{center}
\vspace{-1em}
\end{figure}

We apply the least-square fitting of the linear regression to estimate $\Delta C_D( f_n=1,w_n)$ using the training data $\Delta C^*_D(f_l=1,w_l)$. The fitting curves are plotted in Fig. \ref{fig:fits}.  Since $\Delta C_D(f_n=1,w_n)$ is the decoding complexity reduction of a frame caused by disabling DF of the $n$-th CTU in this frame (i.e., $f_n=1$), its value is also influenced by the total number of CTUs in a frame. For example, in high resolution videos, DF of one CTU occupies less decoding complexity proportion of the whole frame, than that in
lower resolution videos. Such influence can be removed by multiplying $N$, and the function of $N\Delta C_D(f_n=1,w_n)$ for different resolutions can be at the same scale and then trained together. For training $N\Delta C_D(f_n=1,w_n)$, $N\Delta C^*_D(f_l=1,w_l)$ of 3,000 randomly chosen CTUs are used as training samples, and each dot in Fig. \ref{fig:fits} stands for one sample of $(w_l,N\Delta C^*_D(f_l=1,w_l))$. Here, the configurations of the encoder and decoder for training are the same as those in experiments of Section \ref{sec:settings}. Consequently, the function of $\Delta C_D( f_n,w_n)$ is
\setcounter{equation}{17}
\begin{equation}\label{fff1}
\small
\Delta C_D( f_n,w_n)= \frac{1}{N}\cdot (a\cdot w_n+b)\cdot f_n,
\end{equation}
where the values of $a$ and $b$ at different QPs are presented in Table \ref{tab:coc}. Finally, $\Delta C_D( f_n,w_n)$ can be modelled.

\begin{figure}
\centering
\begin{minipage}[b]{0.6\linewidth}
  \centerline{\epsfig{figure=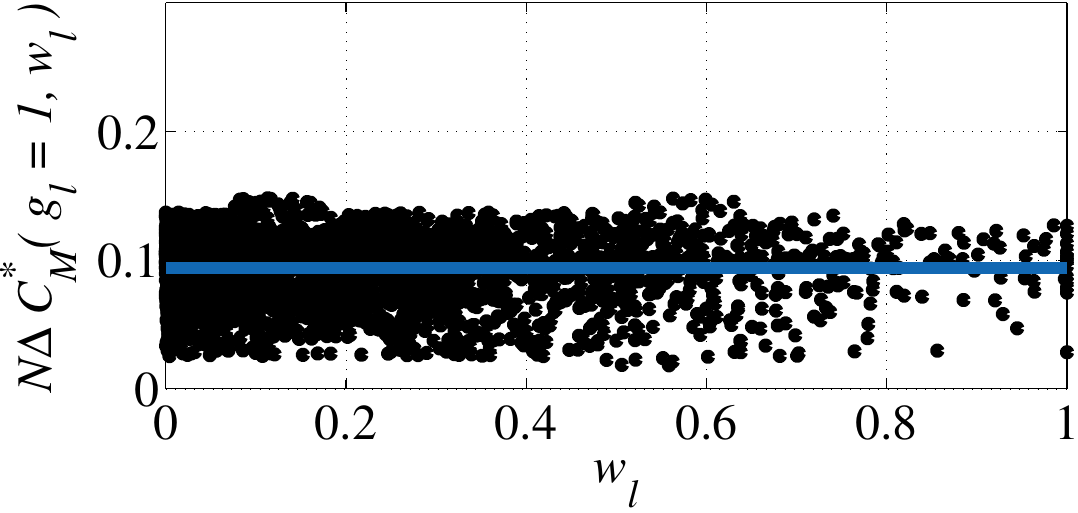,width=5cm}}
\end{minipage}\vspace{-1em} \caption{ \scriptsize{Pairs of $N\Delta C^*_M(g_l=1,w_l)$ versus $w_l$ for QP = 32. }}\label{fig:cmc}
\vspace{-1em}
\end{figure}

\setcounter{equation}{20}
\begin{figure*}[b]
\hrulefill
\begin{equation}
\small
\left\{
\begin{aligned}
  &\min_{{\{f_n\}}_{n=1}^N} \sum_{n=1}^N \Delta  S_D(f_n,w_n) \quad
  \text{s.t.} \quad \sum_{n=1}^N \Delta C_D(f_n,w_n)\geq \Delta C_T,
  \quad\quad\quad\quad\quad\quad\quad\quad\quad\quad \text{if}\ \Delta C_T\leq\sum_{n=1}^N \Delta C_D(f_n=1, w_n),\\
  &\min_{{\{g_n\}}_{n=1}^N} \sum_{n=1}^N \Delta  S_M(g_n,w_n) \quad
  \text{s.t.} \quad \sum_{n=1}^N \Delta C_M(g_n)\geq \Delta C_T-\sum_{n=1}^N\Delta C_D(f_n=1, w_n),
  \quad \text{if}\ \Delta C_T>\sum_{n=1}^N \Delta C_D(f_n=1 ,w_n).
\end{aligned}
\right.\label{rel}
\end{equation}

\begin{equation}
\small\hspace{-4em}
\left\{
\begin{aligned}
  &\min_{{\{f_n\}}_{n=1}^N} \sum_{n=1}^N w_{n}\cdot f_n \quad\quad\quad\quad\quad\quad\quad \text{s.t.} \quad \sum_{n=1}^N \frac{1}{N}\cdot (a\cdot w_n+b)\cdot f_n\geq \Delta C_T, \quad \text{if}\ \Delta C_T\leq\sum_{n=1}^N \frac{1}{N}\cdot(a\cdot w_n+b),\ \text{(a)}\\
   &\min_{{\{g_n\}}_{n=1}^N} \sum_{n=1}^N w_n\cdot(h_1\cdot g_n^3+h_2\cdot g_n^2+h_3\cdot g_n) \quad \text{s.t.} \quad \sum_{n=1}^N \frac{1}{N}\cdot c\cdot g_n \geq \Delta C^{'}_T,
   \quad \text{if}\ \Delta C_T>\sum_{n=1}^N \frac{1}{N}\cdot(a\cdot w_n+b),\ \text{(b)}
\end{aligned}
\right.\label{22}
\end{equation}
\vspace{-1em}
\end{figure*}

\setcounter{equation}{18}

\vspace{-1em}
\subsection{Relationship Modelling for $\Delta C_M(g_n,w_n)$}\label{modelMC}
Similarly, in order to model $\Delta C_M(g_n,w_n)$, four training sequences at four QPs are decoded with MC skipped for $0$, $1$, $2$ and $3$ samples among each four samples (i.e., $g_n=0,1,2,3$). The decoding complexity of each CTU is recorded for all training sequences. Then, we define $\Delta C^*_M(g_l,w_l)$ as the percentage of complexity reduction of a frame, caused by skipping MC of the $l$-th training CTU.

In Fig. \ref{fig:cmc}, we plot the pairs of $N\Delta C^*_M(g_l=1,w_l)$ and $w_l$, when decoding four training sequences at QP = 32. Note that the dots in this figure indicate the pairs of  $(w_l,N\Delta C^*_M(g_l=1,w_l)$ for 3,000 randomly selected CTUs, with the same training configuration as Section \ref{sec:relafn}. Similar results can be found for other values of $g_n$ or other QPs. Generally speaking, this figure indicates that $\Delta C^*_M(g_l,w_l)$ is independent of $w_l$. Therefore, $\Delta C_M(g_n,w_n)$ can be replaced by $\Delta C_M(g_n)$.

\begin{figure}
\vspace{-1em}
\begin{center}
  \subfigure[QP = 22]{\includegraphics[width=.48\linewidth]{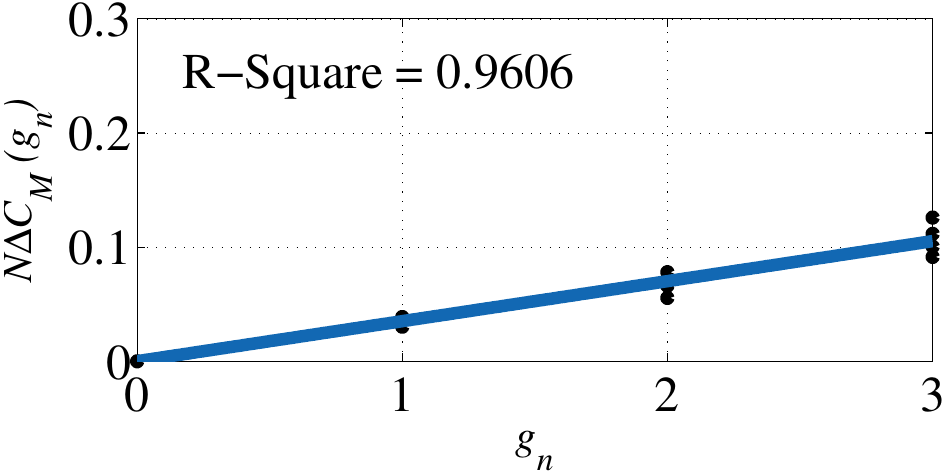}}
  \subfigure[QP = 27]{\includegraphics[width=.48\linewidth]{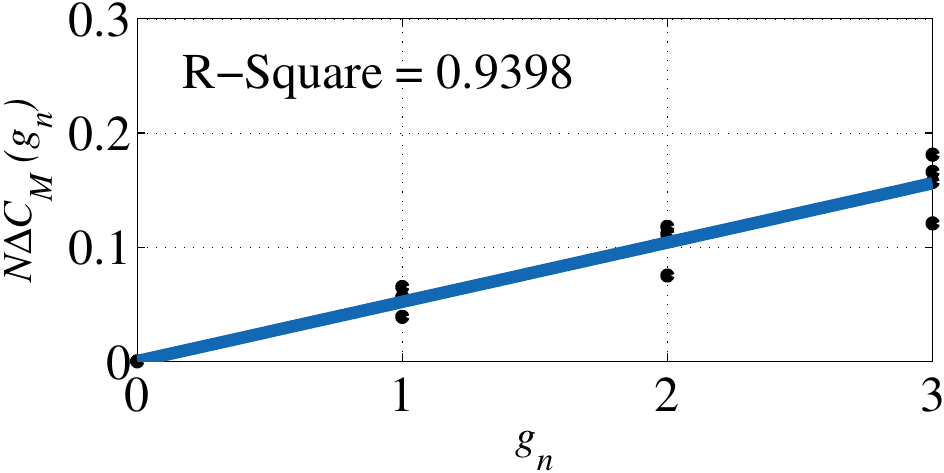}}
  \subfigure[QP = 32]{\includegraphics[width=.48\linewidth]{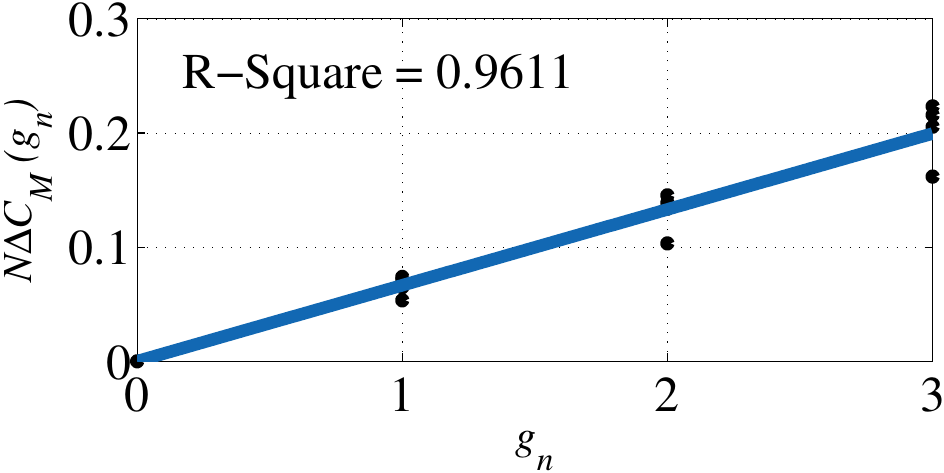}}
  \subfigure[QP = 37]{\includegraphics[width=.48\linewidth]{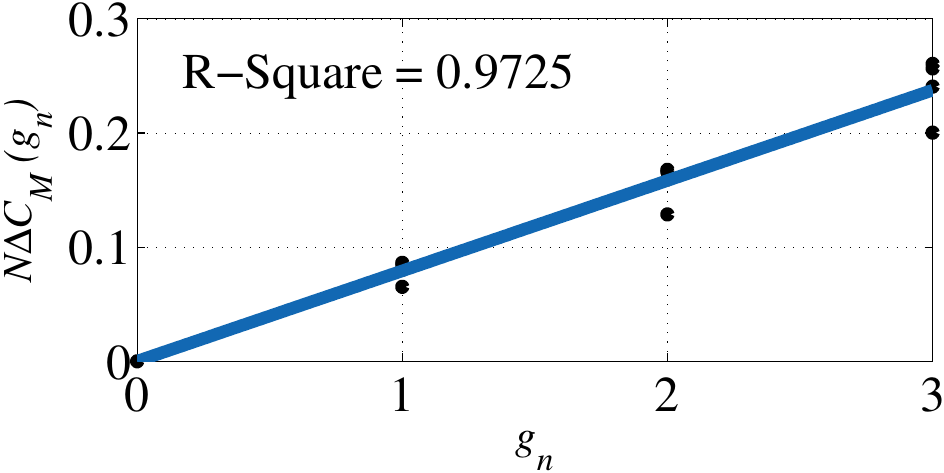}}
  \vspace{-.5em}
\caption{\footnotesize{Fitting curves of $g_n$ versus $N\Delta C_M(g_n)$. Each dot indicates a pair of $(g_n,N\overline{\Delta C^*_M}(g_n))$ where $g_n\in\{0,1,2,3\}$.}}\label{fig:cgnfits}
\end{center}
\vspace{-1em}
\end{figure}

Next, we model $\Delta C_M(g_n)$ by learning from training data of $\{\Delta C^*_M(g_l=g_n)|g_n=0, 1, 2, 3\}$.  Sometimes, the CTU number in each training video may be dramatically different, such that the modeling of $\Delta C_M(g_n,w_n)$ may bias toward some of training video sequences. To avoid such bias, we can estimate the averaged complexity reduction of each training video sequence by
\begin{equation}
\small
\overline{\Delta C^*_M}(g_n) = \frac{1}{L}\sum_{l=1}^{L}\Delta C^*_M(g_l=g_n),
\end{equation}
for each possible value of $g_n$. Recall that $L$ is the total number of CTUs on the training sequences. Then, we have $\overline{\Delta C^*_M}(g_n)$ for each training sequence at a specific QP. For each case of a possible QP value (22, 27, 32 and 37), the least-square fitting of the linear regression is applied on all training data $N\overline{\Delta C^*_M}(g_n)$ of four training sequences. Similar with modelling $\Delta C_D(f_n,w_n)$, we use $N\overline{\Delta C^*_M}(g_n)$ rather than $\overline{\Delta C^*_M}(g_n)$ here to make the regression general for different resolutions. The fitting curves are plotted in Fig. \ref{fig:cgnfits}. Accordingly, the function of $\Delta C_M(g_n)$ is obtained in the following,
\begin{equation}\label{fff2}
\small
\Delta C_M(g_n) = \frac{1}{N}\cdot c\cdot g_n,
\end{equation}
where the values of $c$ at different QPs are presented in Table \ref{tab:coc}. Finally, $\Delta C_M(g_n)$ can be modelled. It is worth pointing out that the training sequences are encoded by hierarchical GOP structure, and the frame-level QP has the offset of $0 \sim +4$ (\textit{encoder\_randomaccess\_main.cfg}). For example, when setting QP = 22, its frame-level QP ranges from 22 to 26. Therefore, in our SGCC approach, the trained parameters $a$,$b$ and $c$ for QP = 22 are to be applied for frames with QP ranging from 22 to 26. Similar setting holds for QP = 27, 32 and 37.

\setcounter{equation}{22}

\section{Solution to SGCC optimization formulation} \label{}
In this section, we concentrate on solving our SGCC formulation of (\ref{f2}), to achieve complexity control of HEVC decoding. Since Fig. \ref{fig:MSE} has shown that the loss of MSE caused by disabling DF is significantly less than that by simplifying MC, there exists $\Delta S_D(f_n,w_n)\ll \Delta S_M(g_n,w_n)$. Therefore, in our SGCC approach, we do not simplify MC when the target complexity $\Delta C_T$ can be achieved by only disabling DF, i.e., $\Delta C_T\leq\sum_{n=1}^N \Delta C_D(f_n=1 ,w_n)$). The MC is simplified only if $\Delta C_T$ cannot be satisfied by disabling DF, i.e., $\Delta C_T>\sum_{n=1}^N \Delta C_D(f_n=1 ,w_n)$). As such, we can rewrite (\ref{f2}) of our SGCC formulation as \eqref{rel}. Moreover, as shown in \eqref{rel}, the perceptual quality loss is minimized by the optimization term, in which the DF and MC of non-salient CTUs are disabled/simplified in priority.

As discussed in Section \uppercase\expandafter{\romannumeral3}-A, we replace $\Delta S_D(f_n,w_n)$ and $\Delta S_M(g_n,w_n)$ of \eqref{rel} by their normalized functions $\Delta \widetilde S_D(f_n,w_n)$ and $\Delta \widetilde S_M(g_n,w_n)$. Then, given the relationship of (\ref{sd_nor}), (\ref{sm_nor}), (\ref{fff1}) and (\ref{fff2}), formulation (\ref{rel}) can be finally turned to \eqref{22}, where $\Delta C^{'}_T = \Delta C_T-\sum_{n=1}^N \frac{1}{N}\cdot(a\cdot w_n+b)$.

\begin{table}[!t]
\vspace{-1em}
\begin{center}
\scriptsize
\caption{\footnotesize{Parameters in relationship modelling for our SGCC approach.}} \label{tab:coc}
\begin{tabular}{|c|c|c|c|c|}
  \hline
        & QP = 22 & QP = 27 & QP = 32 & QP = 37\\
  \hline
  $h_1$ & \multicolumn{4}{c|}{0.1040}\\
  \hline
  $h_2$ & \multicolumn{4}{c|}{-0.2737}\\
  \hline
  $h_3$ & \multicolumn{4}{c|}{0.2184}\\
  \hline
  $a$& 0.3041 &0.3874&0.4101& 0.4347 \\
  \hline
  $b$& 0.0255&0.0433&0.0459& 0.0576\\
  \hline
  $c$& 0.0351 &0.0520&0.0665& 0.0792 \\
  \hline
\end{tabular}
\end{center}
\vspace{-2em}
\end{table}

\begin{figure}[t]
\vspace{-1em}
\begin{center}
  \subfigure{\includegraphics[width=0.97\linewidth]{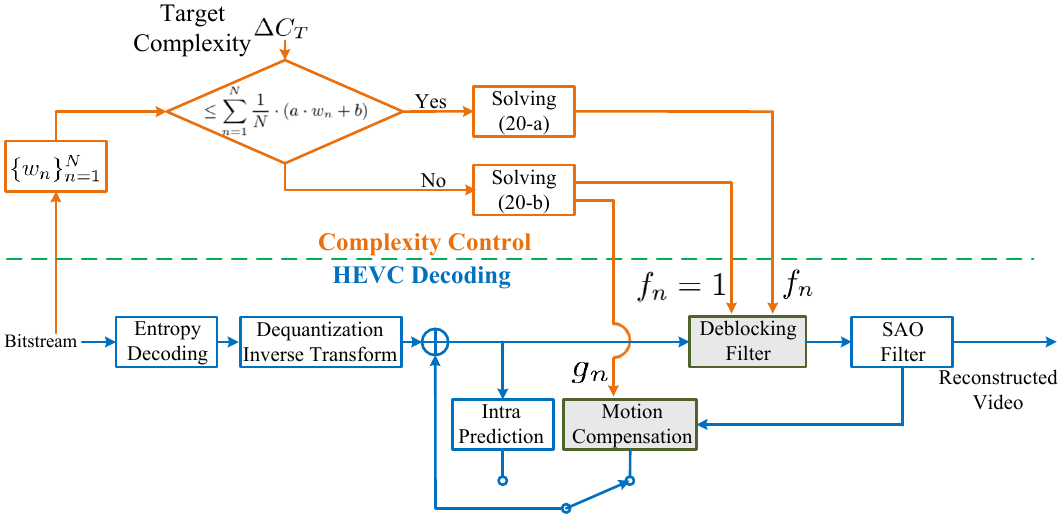}}
 \vspace{-.5em}
\caption{\footnotesize{Framework of our SGCC approach.}}\label{fig:sgcc}
\end{center}
\vspace{-2em}
\end{figure}

Given the above equations, we only need to solve (22-a) when the target complexity $\Delta C_T\leq\sum_{n=1}^N \frac{1}{N}\cdot(a\cdot w_n+b)$. When $\Delta C_T>\sum_{n=1}^N \frac{1}{N}\cdot(a\cdot w_n+b)$, we need to solve (22-b) with DF of all CTUs disabled. Once (22-a) and (22-b) are solved, the decoding complexity of HEVC can be reduced to the target by our SGCC approach, as summarized in Fig. \ref{fig:sgcc}. As seen from this figure, before decoding each CTU, our SGCC approach decides how to simplify MC and whether to enable DF, without any change on the CTU-level decoding pipeline. Next, we discuss how to solve (22-a) and (22-b), respectively.

\vspace{-1em}
\subsection{Solution to formulation (22-a)}
First, we aim at finding optimal solution $\textbf{F} = \{ f_n\}_{n=1}^{N}$ of (22-a). First, let $\{\widetilde{w}_n\}_{n=1}^{N}$ be the set of the ascending sorted $\{w_n\}_{n=1}^{N}$. Given $\{\widetilde{w}_n\}_{n=1}^{N}$, Lemma 3 can be used for finding the optimal solution to (22-a).

\begin{lemma} Let $a>0$, $b>0$, $l>0$ and $w_n\in [0,1]$. Assume that $\textbf{F} = \{ f_n\}_{n=1}^{N}$ satisfies
\begin{equation} \label{fnn}
\small
 f_n=\left\{
\begin{aligned}
1&,& \quad w_n \leq \widetilde{w}_I \\
0&,& \quad \text{otherwise}, \\
\end{aligned}
\right.
\end{equation}
where $\tilde{w}_I$ is the $I$-th value of ascending sorted $\{w_n\}_{n=1}^{N}$.
Assume that $\textbf{F}^{'} = \{f^{'}_n\}_{n=1}^{N}$ is another set with $f^{'}_n\in\{0,1\}$.

If
\begin{equation} \label{case}
\small
 \sum_{n=1}^{N} \frac {1}{N} (a\cdot w_n +b) \cdot f_n = \sum_{n=1}^{N} \frac {1}{N} (a\cdot w_n +b) \cdot f_n^{'},
\end{equation}
then the following inequality holds
\begin{equation} \label{result}
\small
\sum_{n=1}^{N}w_n\cdot f_n \leq \sum_{n=1}^{N}  w_n\cdot f_n^{'}.
\end{equation}
\end{lemma}
\begin{Proof} The proof for Lemma 3 is in Appendix A.  \hfill{$\blacksquare$}

\end{Proof}

According to Lemma 3, if and only if $w_n \leq \widetilde{w}_I$, $ f_n=1$ is the optimal solution to (\ref{22}-a). In order to minimize $\sum_{n=1}^N w_n\cdot f_n$ at the constraint of $\sum_{n=1}^N \frac{1}{N}(a\cdot w_n + b)\cdot f_n \geq \Delta C_T$, $\sum_{n=1}^N \frac{1}{N}(a\cdot w_n + b)\cdot f_n$ should be as close to $\Delta C_T$ as possible. Consequently, the optimal solution to (22-a) can be obtained as
\begin{equation}
\small
 f_n=\left\{
\begin{aligned}
1&,& \quad w_n \leq \widetilde{w}_I \\
0&,& \quad \text{otherwise}, \\
\end{aligned}
\right. \label{fn}
\end{equation}
where $I$ satisfies
\begin{equation}
\small
\frac{1}{N}\sum_{n=1}^I( a\cdot\widetilde{w}_n+b) \geq \Delta C_T > \frac{1}{N}\sum_{n=1}^{I-1} (a\cdot \widetilde{w}_n+b).
\label{I}
\end{equation}
In our SGCC approach, the solution of $I$ to \eqref{I} is searched by the following way. For each frame, $a\cdot\widetilde{w}_n+b$ is calculated starting from $n=1$, and then added up for $n=1,2,\ldots,I$ until its sum is  $\geq N\cdot \Delta C_T$. Once the sum $\sum_{n=1}^I( a\cdot\widetilde{w}_n+b)$ $\geq N\cdot \Delta C_T$, a suitable $I$ can be found out.

\subsection{Solution to formulation (22-b)}
Next, we discuss on the solution to formulation (22-b). First, (22-b) can be simplified by Lemma 4.

\begin{lemma} The nonlinear integer programming (22-b) is equivalent to the linear integer programming problem as follows,
\begin{equation}
\small
\begin{split}\label{22-b-2}
&\vspace{-1em}\min_{N_3,N_2,N_1}\!\!\!\sum_{n=1}^{N_3}\!\!\widetilde w_n\!+\!\!\!\!\!\!\!\sum_{n=N_3\!+\!1}^{N_3\!+\!N_2}\!\!\!\!\!(8h_1\!+\!4h_2\!+\!2h_3)\!\cdot\!\widetilde w_n\!\!+\!\!\!\!\!\!\!\!\!\!\!\sum_{n=N_3\!+\!N_2\!+\!1}^{N_3\!+\!N_2\!+\!N_1}\!\!\!\!\!\!\!\!\!(h_1\!+\!h_2\!+\!h_3)\!\cdot\!\widetilde w_n  \\
&\text{s.t.} \quad \frac{1}{N}\cdot c \cdot (N_1\!+\!2 N_2\!+\!3 N_3) \geq \Delta C_T^{'}.
\end{split}
\end{equation}
In \eqref{22-b-2}, $N_1$, $N_2$ and $N_3$ are the numbers of CTUs corresponding to $g_n=1,2$ and $3$ in a frame, and they satisfy $N_1+N_2+N_3\leq N$.
\end{lemma}
\begin{Proof} The proof for Lemma 4 is in Appendix B.  \hfill{$\blacksquare$}
\end{Proof}

According to Lemma 4, the optimal solution to (22-b) can be obtained, once the formulation of \eqref{22-b-2} is worked out. In fact, \eqref{22-b-2} is a linear programming problem, which can be solved by the branch-and-bound algorithm \cite{li2006nonlinear}. However, the computational complexity of the solution is still enormous, especially for large CTU number $N$ in a frame with high resolution. It is because the branch-and-bound algorithm has to be carried out to solve \eqref{22-b-2} for each frame. Next, we further simplify \eqref{22-b-2} to reduce its computational complexity.

\begin{proposition}$\widetilde w_n$ is of almost uniform distribution as follows,
\begin{equation}\label{sumw}
\small
\forall N_t \in \{n\}_{n=1}^N, \quad \sum_{n=1}^{N_t}\widetilde w_n \approx k\cdot N_t^2,
\end{equation}
where $k$ is a positive constant.
\end{proposition}
\begin{Proof} The proof for Proposition 5 is in Appendix C. \hfill{$\blacksquare$}
\end{Proof}

Based on Proposition 5, \eqref{22-b-2} can be rewritten by
\begin{equation}\label{22-b-3}
\small
\begin{aligned}
\min_{N_3,N_2,N_1} N_3^2+(8h_1+4h_2+2h_3)\cdot ((N_2+N_3)^2-N_3^2)\\
+(h_1+h_2+h_3)\cdot((N_1+N_2+N_3)^2-(N_2+N_3)^2) \\
\text{s.t.}\quad \frac{1}{N}\cdot c \cdot (N_1+2 N_2+3 N_3)\geq \Delta C_T^{'}.
\end{aligned}
\end{equation}
Note that $k$ is a constant which is independent of the minimization problem in \eqref{22-b-3}, and thus $k$ can be simply removed from the minimization formulation.

Next, we apply the branch-and-bound algorithm \cite{li2006nonlinear} to solve \eqref{22-b-3}, and it only needs to be solved once before decoding. We establish a table for the solutions to \eqref{22-b-3} at each specific $\Delta C_T^{'}$. Then, given $\Delta C_T^{'}$, we can simply obtain $N_3, N_2$ and $N_1$ by table look-up. This way, the overhead of computational complexity on solving (22-b) can be avoided.

\begin{figure}
%\vspace{-1em}
\begin{center}
 % \subfigure{\includegraphics[width=.12\linewidth]{legend1}}
  \subfigure{\includegraphics[width=.98\linewidth]{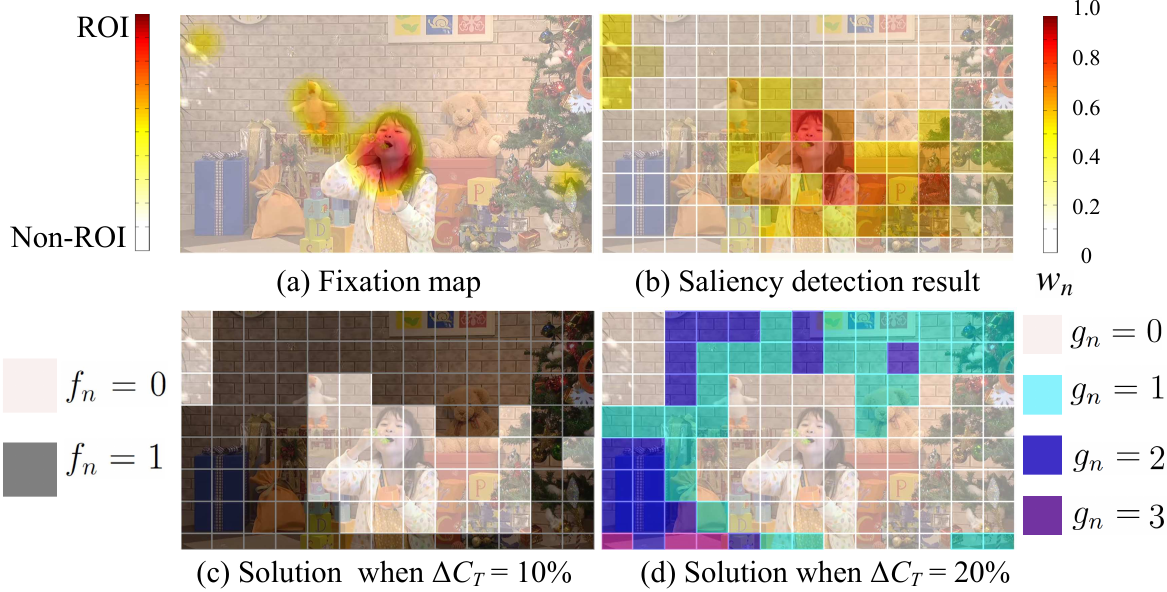}}
 % \subfigure{\includegraphics[width=.12\linewidth]{legend}}
  \vspace{-.5em}
\caption{\footnotesize{(a) is the fixation map (groundtruth ROI) of frame 274 in \textit{PartyScene}, which is obtained from the database of \cite{xu2017learning}. (b) is the saliency detection results. (c) and (d) are example solutions of $\{f_n\}_{n=1}^n$ and $\{g_n\}_{n=1}^n$, when $\Delta C_T = 10\%$ and $20\%$, respectively. Note that (d) only shows the values of $g_n$, in which the values of $f_n$ are all equivalent to 1. }}\label{fig:f_ng_n}
\end{center}
\vspace{-1.5em}
\end{figure}

An example of $\{f_n\}_{n=1}^N$ and $\{g_n\}_{n=1}^N$ solved by our SGCC approach is shown in Fig. \ref{fig:f_ng_n}-(c) and (d). As can be seen , larger $f_n$ or $g_n$ corresponds to smaller $w_n$, which is the saliency value as illustrated in Fig. \ref{fig:f_ng_n}-(b). The detected saliency map, shown in Fig. \ref{fig:f_ng_n}-(b), tallies well with the groundtruth, i.e., the fixation map in Fig. \ref{fig:f_ng_n}-(a). As a result, the decoding complexity of CTUs in non-ROI is reduced in high priority, and the quality of ROI (e.g., face) rarely degrades. This indicates that the perceptual quality loss can be minimized in applying our SGCC approach.

\vspace{-1em}
\subsection{Error propagation analysis}\label{error}

\begin{figure}[!t]
\begin{center}
  \subfigure{\includegraphics[width=.85\linewidth]{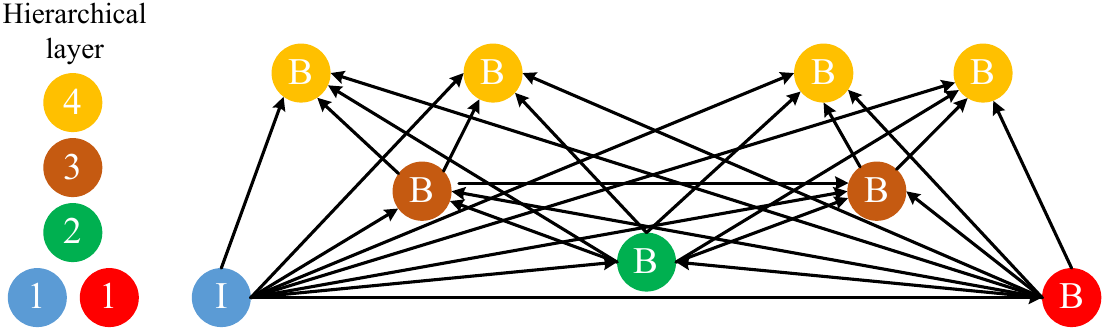}}
 \vspace{-1em}
\caption{\footnotesize{GOP structure and its hierarchical layers.}}\label{layer}
\vspace{-1em}
\end{center}
\end{figure}

The quality loss of each decoded frame, which is caused by the above complexity control, may propagate to other frames predicted by this frame. Hence, it is necessary to analyze the error propagation across decoded frames. We find through the following observations that the hierarchical coding structure of HEVC can significantly alleviate the error propagation in our SGCC approach. Here, for analysis, we use the hierarchical GOP structure of the default HM Random Access (RA) with \emph{encoder\_randomaccess\_main.cfg} file, as shown in Fig. \ref{layer}. Similar results can be found for other GOP structure.

\begin{figure}[!t]
\begin{center}
  \subfigure{\includegraphics[width=.8\linewidth]{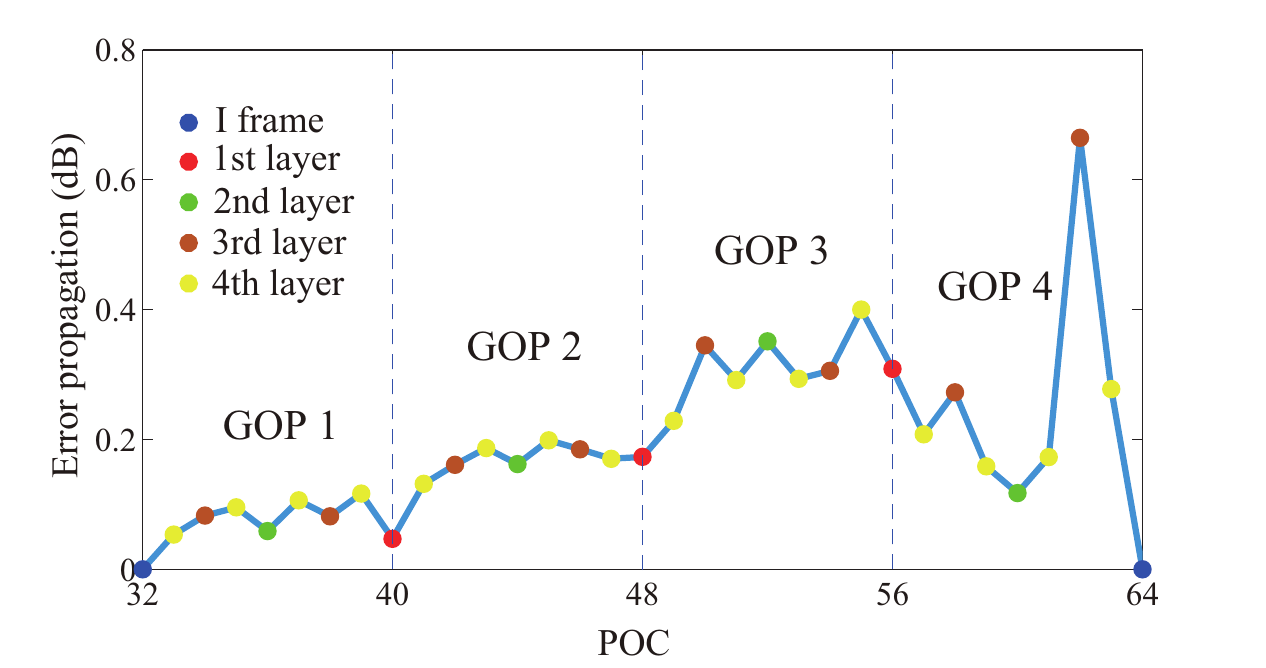}}
 \vspace{-1em}
\caption{\footnotesize{Averaged propagation error of each frame at QP = 32 and $\Delta C_T = 20\%$, in terms PSNR reduction, along with Picture Order Count (POC).}}\label{propagation_error}
\end{center}
 \vspace{-1em}
\end{figure}

\begin{observation}\label{ob1}
The quality loss of I frames does not incur any error propagation, when reducing decoding complexity by our SGCC approach.
\end{observation}
\setcounter{Analysis}{5}
\begin{Analysis}
The reconstruction of I-frames is independent of other frames, and thus the quality loss of other frames has no impact on each decoded I frames. We further tested the error propagation of two neighboring I frames and four GOP between them (from frame 32 to frame 64), averaged over four training sequences. Here, the error propagation of the $i$-th frame is calculated as follows. First, we only apply our SGCC approach on frame $i$, and do not make any complexity reduction on other frames. Then, the quality of the $i$-th frame is evaluated by Y-PSNR in dB. For the anchor, we apply our SGCC approach on all frames, and also measure the quality of the $i$-th frame by Y-PSNR. Finally, the difference of above two PSNRs is calculated as the error propagation at frame $i$. The results are shown in Fig. \ref{propagation_error}, and we find that the PSNR reduction of each I frame is 0 dB.

Additionally, the quality loss of I frames does not incur any error propagation within the frame for our SGCC approach, as only intra prediction mode is applied in I frames. In the I-frames decoding, the DF is implemented in every frame after the reconstruction (intra prediction, etc.) of the whole frame \cite{sullivan2012overview}. Thus, the quality degradation caused by disabling DF cannot propagate through intra prediction among CTUs.
Furthermore, since MC is only related to inter frame prediction, the error cannot propagate within I frame. This completes the validation of Observation 6.
\end{Analysis}

\begin{observation}\label{ob2}
The quality loss of B or P frames incurs small error propagation due to the hierarchical coding structure in HEVC, when reducing decoding complexity by our SGCC approach.
\end{observation}
\begin{Analysis}
In B or P frames, because of inter prediction, the quality degradation of the reference frames is possible to propagate to the currently decoded frame. Thus, error propagation exists in B or P frames. However, the error propagation is restricted to be small by the hierarchical GOP structure. First, each I frame does not incur any error propagation as pointed out by Observation \ref{ob1}. In addition, I frames do not have MC, such that their quality loss is only from disabling DF, which is significantly lower than that of simplifying MC (see Fig. \ref{fig:MSE}). As a result,  after I frames, the error propagation of B or P frames terminates, and their quality loss is resumed to be small. Second, although the B or P frames, especially in higher layers or far from I frames, suffer from error propagation, the reference frames at different layers of hierarchical coding structure ensure (see Fig. \ref{layer}) that each decoded frame is predicted by several frames. Specifically, all frames of the first GOP after I frames are all predicted by I frames, which has little quality loss. Then, for the second GOP all frames have the reference frame directly predicted by I frames, such that the shortest prediction path to I frames is one frame. The shortest prediction path to I frames is two frames for the third GOP, and so on. Note that the error propagation of the frames of the GOP before an I frame can be reduced to be small, as they are also predicted by the incoming I frame. Therefore, in the hierarchical coding structure of HEVC, there exists small error propagation for the quality loss of B/P frames.

In addition, Fig. \ref{propagation_error} shows the error propagation of all B frames between two neighboring frames averaged over four training sequences, when $\Delta C_T = 20\%$ and QP = 32. As shown, the averaged error propagation of B frames is only 0.19 dB. Thereby, we can conclude that the error propagation of quality loss for B or P frames is rather small. Finally, the analysis of Observation 7 is completed.
\end{Analysis}

Note that when applying our SGCC approach, I frames should be inserted to terminate error propagation, according to Observation 6. In this paper, the period between I frames is set as 32 for training and test sequences, according to the default configuration of \emph{encoder\_randomaccess\_main.cfg}.

\begin{table*}[!t]
  \footnotesize
  \centering
  \caption{Individual MAR contributions  ($\%$) of DF disabling and MC simplification.}
    \begin{tabular}{|c|r|c|c|c|c|c|c|c|c|c|c|c|c|}
    \hline
    \multirow{2}[4]{*}{\textbf{Classes}} & \multicolumn{1}{c|}{\multirow{2}[4]{*}{\textbf{Sequences}}} & \multicolumn{3}{c|}{MAR of QP = 22} & \multicolumn{3}{c|}{MAR of QP = 27} & \multicolumn{3}{c|}{MAR of QP = 32} & \multicolumn{3}{c|}{MAR of QP = 37} \\
\cline{3-14}          & \multicolumn{1}{c|}{} & DF    & MC    & SGCC   & DF    & MC    & SGCC   & DF    & MC    & SGCC   & DF    & MC    & SGCC \\
    \hline
    \multirow{2}[2]{*}{A} &\textit{Traffic} & 13.98 & 19.20 & 33.18 & 14.67  & 25.12  & 39.79  & 14.93  & 27.03  & 41.96  & 14.45  & 30.19  & 44.64  \\
\cline{2-14}          & \textit{PeopleOnStreet} & 15.57  & 9.60  & 25.17  & 19.58  & 11.55  & 31.13  & 21.78  & 12.63  & 34.41  & 20.61  & 17.91  & 38.52  \\
    \hline
    \multirow{3}[2]{*}{B} & \textit{Kimono} & 12.90  & 15.35  & 28.25  & 14.18  & 22.93  & 37.11  & 14.49  & 25.71  & 40.20  & 10.80  & 31.07  & 41.87  \\
\cline{2-14}          & \textit{ParkScene} & 13.47  & 17.12  & 30.59  & 14.33  & 22.47  & 36.80  & 14.89  & 25.21  & 40.10  & 11.55  & 32.39  & 43.94  \\
\cline{2-14}          & \textit{BQTerrace} & 10.98  & 12.66  & 23.64  & 16.18  & 19.84  & 36.02  & 16.14  & 25.88  & 42.02  & 12.83  & 32.40  & 45.23  \\
    \hline
    \multirow{2}[2]{*}{C} & \textit{RaceHorses} & 12.54  & 6.29  & 18.83  & 15.92  & 11.03  & 26.95  & 15.30  & 16.56  & 31.86  & 16.46  & 17.93  & 34.39  \\
\cline{2-14}          & \textit{PartyScene} & 10.67  & 9.71  & 20.38  & 15.81  & 13.28  & 29.09  & 14.28  & 20.73  & 35.01  & 11.27  & 26.74  & 38.01  \\
    \hline
    \multirow{4}[2]{*}{D} & \textit{RaceHorses} & 13.19  & 8.45  & 21.64  & 16.38  & 10.08  & 26.46  & 18.76  & 10.77  & 29.53  & 16.78  & 17.97  & 34.75  \\
\cline{2-14}          & \textit{BQSquare} & 10.72  & 14.60  & 25.32  & 12.36  & 20.49  & 32.85  & 13.70  & 23.55  & 37.25  & 12.48  & 28.34  & 40.82  \\
\cline{2-14}          & \textit{BlowingBubbles} & 10.62  & 12.85  & 23.47  & 12.61  & 16.47  & 29.08  & 14.02  & 18.68  & 32.70  & 15.36  & 21.50  & 36.86  \\
\cline{2-14}          & \textit{BasketballPass} & 12.94  & 10.21  & 23.15  & 15.98  & 12.51  & 28.49  & 17.73  & 13.78  & 31.51  & 15.12  & 20.41  & 35.53  \\
    \hline
    \multicolumn{2}{|c|}{\textbf{Average}} & \textbf{12.51 } & \textbf{12.37 } & \textbf{24.87 } & \textbf{15.27 } & \textbf{16.89 } & \textbf{32.16 } & \textbf{16.00 } & \textbf{20.05 } & \textbf{36.05 } & \textbf{14.34 } & \textbf{25.17 } & \textbf{39.51 } \\
    \hline
    \end{tabular}%
  \label{tab:contr}%
\end{table*}%

\vspace{-1em}
\subsection{Complexity overhead analysis}\label{overhead}
Finally, we analyze the complexity overhead in applying our SGCC approach.
The complexity overhead of our SGCC approach includes calculating $\{w_n\}$, computation on (22-a) and (22-b). Their computational time is evaluated and reported in Table \ref{tab:overhead}. Note that the function \textit{QueryPerformanceCounter}() in Visual C++ was used to record the computational time. The experiment was performed on a Windows PC with Inter(R) Core(TM) i7-4790K CPU.

It can be seen from Table \ref{tab:overhead} that the complexity overhead of our SGCC approach is rather small. In particular, calculating saliency values $\{w_n\}$ consumes averagely 0.058 ms per 1080p frame. When calculating (22-a), saliency values $\{w_n\}_{n=1}^{N}$ need to be sorted as $\{\widetilde{w}_n\}_{n=1}^{N}$ by the quicksort algorithm, which averagely consumes 0.010 ms per 1080p frame. Besides, computing $I$ in \eqref{I} consumes averagely 0.001 ms per frame for 1080p videos for solving (22-a). For solving (22)-b, as mentioned in Section IV-B, we establish a look-up table for the solutions to (22)-b, and we can simply obtain the solution by the table look-up, when decoding HEVC bitstreams. Therefore, the computational time of (22)-b is only for reading the values $N_1$, $N_2$ and $N_3$ from the table, according to given $\Delta C_T$. Such computational time is too little to be measured. In summary, the total complexity overhead of decoding complexity control for 1080p sequences is 0.069 ms per frame, which is very little compared to DF and MC in HEVC decoding. For other resolutions, similar computational time can be found in Table \ref{tab:overhead}.

Note that when applying our SGCC approach to the HEVC bitstreams with other configurations, the parameters in Table \ref{tab:coc} may need to be re-trained. However, these parameters only need to be re-trained once before applying the SGCC approach to the HEVC decoder. Consequently, re-training the parameters only introduces off-line computational time overhead, and does not consume any HEVC decoding time. In our SGCC approach, training a new set of parameters consumes around 297.40 s overhead of computational time in off-line manner.

\begin{table}[!t]
  \centering
  \footnotesize
  \caption{Complexity overhead of our SGCC approach per frame.}
    \begin{tabular}{|c|c|c|c|c|}
    \hline
     & Calculating $\{w_n\}_{n=1}^N$ & Solving (22-a) & (22-b) & Total \\
    \hline
    1600p & 0.119 ms & 0.025 ms  & -     & 0.144 ms \\
    \hline
    1080p & 0.058 ms & 0.011 ms & -     & 0.069 ms  \\
    \hline
    480p & 0.015 ms & 0.002 ms  & -     & 0.017 ms \\
    \hline
    240p & 0.003 ms &  0.001 ms & -     & 0.004 ms \\
    \hline
    \end{tabular}%
   \vspace{-1em}
  \label{tab:overhead}%
\end{table}%

\section{Experimental Results}
In this section, experimental results are presented to validate the effectiveness of our SGCC approach, in comparison with the latest HEVC decoding complexity reduction approaches \cite{nogues2014power} and \cite{nogues2015modified}.

\vspace{-1em}
\subsection{Settings}\label{sec:settings}
All 15 sequences of Classes A-D (except 10-bit sequences) from the JCT-VC database \cite{ohm2012comparison} were divided into non-overlapping training and test sets. Four sequences were selected as the training set to learn the relationship of Section \uppercase\expandafter{\romannumeral3}. Then, we tested our approach on the remaining sequences, including two $2560\times1600$ sequences \emph{Traffic} and \emph{PeopleOnStreet} from Class A, three $1920\times1080$ sequences \emph{Kimono}, \emph{ParkScene} and \emph{BQTerrace} from Class B, two $832\times480$ sequences \emph{RaceHorses} and \emph{PartyScene} from Class C, and four $416\times240$ sequences \emph{RaceHorses}, \emph{BQSquare}, \emph{BlowingBubbles} and \emph{BasketballPass} from Class D. First, all tested sequences were encoded by the HM 16.0 encoder. Here, the configuration of RA was implemented with GOP size being 8. Four common QPs, i.e., 22, 27, 32 and 37, were chosen to encode the test sequences. All other parameters were set by default in the encoder, using the \emph{encoder\_randomaccess\_main.cfg} file. Besides, HM 16.0 with its default settings was also utilized as the decoder. In our experiments, our SGCC approach is implemented in the HM platform, the same as most of existing HEVC complexity reduction works \cite{tan2016fast,vanne2014efficient,pan2016fast,deng2014complexity,zhang2015machine,deng2015subjective}. Compared to encoding, HM is more practical in decoding, since our experiments found that it is able to achieve real-time decoding for 1080p videos at 24 fps and QP = 37 on a Windows PC with Inter(R) Core(TM) i7-4790K CPU.

The experiments were all performed on a Windows PC with Inter(R) Core(TM) i7-4790K CPU and 32G RAM. To evaluate visual quality, both Y-PSNR difference ($\Delta \text{PSNR}$) and Eye-tracking Weighted Y-PSNR difference ($\Delta\text{EW-PSNR}$) \cite{li2011visual} are assessed. Here, Y-PSNR and EW-PSNR are calculated upon the raw and decoded sequences. Then, $\Delta \text{PSNR}$ and $\Delta\text{EW-PSNR}$ quantify the PSNR and EW-PSNR degradation, when decoding sequences by HEVC with our SGCC, \cite{nogues2014power} and \cite{nogues2015modified} approaches, instead of the original HEVC decoder. As such, the smaller $\Delta \text{PSNR}$ and $\Delta\text{EW-PSNR}$ indicate better performance in quality loss. In calculating $\Delta\text{EW-PSNR}$, we utilize human fixation maps from eye-tracking experiment to weight MSE, for fair comparison. In addition, the results of the Difference Mean Opinion Score (DMOS) \cite{recommendation2002500} are also measured to assess the subjective quality of decoding sequences.
\begin{figure}
\centering
  \subfigure{\includegraphics[width=.8\linewidth]{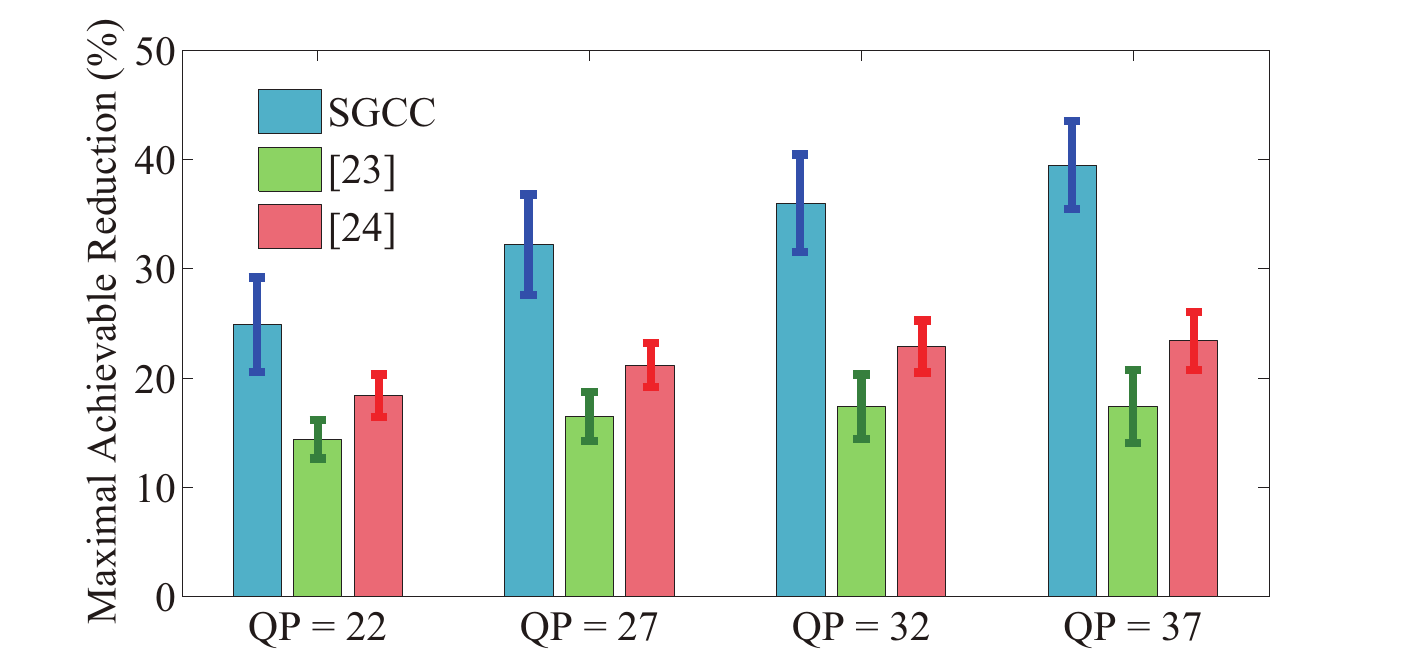}}\\
 \vspace{-1em}
\caption{\footnotesize{Mean and standard deviation of MARs for our SGCC, [23] and [24] approaches.}}\label{fig:reduction}
\vspace{-2em}
\end{figure}

\begin{table*}[!t]
\vspace{-2em}
\caption{\footnotesize{Complexity control error of our SGCC approach.}} \label{tab:complexity}
\vspace{-1em}
\centering
\scriptsize
\begin{tabular}{|c|c|c|c|c|c|c|c|c|c|c|c|c|c|}
\hline
\multirow{3}[6]{*}{\textbf{Classes}} & \multirow{3}[6]{*}{\textbf{Sequences}} & \multicolumn{2}{c|}{QP = 22} & \multicolumn{3}{c|}{QP = 27} & \multicolumn{3}{c|}{QP = 32} & \multicolumn{4}{c|}{QP = 37} \\
\cline{3-14}      &       & \multicolumn{2}{c|}{$\Delta C_T$ (\%)} & \multicolumn{3}{c|}{$\Delta C_T$ (\%)} & \multicolumn{3}{c|}{$\Delta C_T$ (\%)} & \multicolumn{4}{c|}{$\Delta C_T$ (\%)} \\
\cline{3-14}      &       & \textbf{10} & \textbf{20} & \textbf{10} & \textbf{20} & \textbf{30} & \textbf{10} & \textbf{20} & \textbf{30}  & \textbf{10} & \textbf{20} & \textbf{30} & \textbf{40} \\
\hline
\multirow{2}[4]{*}{A} & \textit{Traffic} & +2.81  & +4.46  & +0.97  & +3.01  & +7.30  & -0.19  & +0.78  & +4.14    & +0.44  & -0.29  & +2.20  & +4.29  \\
\cline{2-14}      & \textit{PeopleOnStreet} & +3.06  & -0.02  & +2.10  &+3.20  & +1.13  & +0.83  & +5.06  & +0.89   & +2.34  & +5.48  & +2.12  & -1.72  \\
\hline
\multirow{3}[6]{*}{B} & \textit{ParkScene} & +1.82  & +2.54  & +0.57  & +2.16  & +7.11  & +0.47  & +0.87  & +3.03    & +1.78  & +0.08  & +2.05  & +3.48  \\
\cline{2-14}      & \textit{BQTerrace} & -1.70  & -3.02  & +1.57  & +0.94  & +6.80  & +0.50  & -0.45  & +3.67    & +1.62  & -1.43  & +1.89  & +4.81  \\
\cline{2-14}      & \textit{Kimono} & -0.72  & +2.14  & +1.47  & +1.33  & +6.02  & +1.01  & +1.23  & +3.35  & +0.60  & -0.94  & +0.38  & +1.34  \\
\hline
\multirow{2}[4]{*}{C} & \textit{RaceHorses} & +0.39  & -3.93  & +0.29  & -1.14  & -3.05  & -0.40  & +0.83  & -3.83    & +3.23  & +2.27  & -1.77  & -6.09  \\
\cline{2-14}      & \textit{PartyScene} & -1.78  & -3.79  & -1.21  & -2.15  & -0.91  & -2.41  & -1.80  & -2.63    & -0.57  & -1.49  & -1.79  & -2.46  \\
\hline
\multirow{4}[8]{*}{D} & \textit{RaceHorses} & -0.52  & -3.45  & -0.58  & -1.26  & -3.54  & -0.37  & 1.28  & -3.56   & +1.74  & +2.41  & -1.44  & -5.38  \\
\cline{2-14}      & \textit{BQSquare} & -2.74  & -2.15  & -3.31  & -1.20  & +2.85  & -3.14  & -2.40  & -0.14   & -1.91  & -3.22  & -1.40  & +0.24  \\
\cline{2-14}      & \textit{BlowingBubbles} & -3.31  & -3.45  & -3.52  & -2.57  & -0.92  & -3.14  & -2.25  & -2.98    & -1.74  & -2.37  & -2.82  & -3.20  \\
\cline{2-14}      & \textit{BasketballPass} & -0.69  & -2.44  & -1.26  & -0.53  & -1.51  & -1.27  & +0.45  & -2.77   & +0.15  & +0.96  & -1.83  & -4.56  \\
\hline
\multicolumn{2}{|c|}{\textbf{MAE}} & \textbf{1.78 } & \textbf{2.85 } & \textbf{1.53 } & \textbf{1.77 } & \textbf{1.94 } & \textbf{1.25 } & \textbf{1.58 } & \textbf{2.82 }  & \textbf{1.46 } & \textbf{1.90 } & \textbf{1.79 } & \textbf{3.42 } \\
\hline
\multicolumn{2}{|c|}{\textbf{MRE}} & \textbf{17.78 } & \textbf{14.26 } & \textbf{15.32 } & \textbf{8.86 } & \textbf{6.45 } & \textbf{12.47 } & \textbf{7.91 } & \textbf{9.39 }  & \textbf{14.60 } & \textbf{9.50 } & \textbf{5.97 } & \textbf{8.55 } \\
\hline
\end{tabular}%
\vspace{-1em}
\end{table*}

\begin{table*}
\caption{\footnotesize{$\Delta \text{PSNR}$ and $\Delta \text{EW-PSNR}$ (dB) at QP = 32 and 22 for SGCC, \cite{nogues2014power} and \cite{nogues2015modified}.}} \label{tab:psnr&w-psnr}
\vspace{-1em}
\centering
\scriptsize
\begin{tabular}{|p{.42cm}<{\centering}|p{1.42cm}<{\centering}|p{.5cm}<{\centering} |p{1.92cm}<{\centering}|c|p{1.97cm}<{\centering}|c|p{2.05cm}<{\centering}|c|}
\hline
\multirow{2}[2]{*}{\textbf{Class}} & \multirow{2}[2]{*}{\textbf{Sequence}}& \multirow{2}[2]{*}{\textbf{Appr.}} & \multicolumn{2}{c|}{QP=32, $\Delta\!C_T\!\!\!=\!\!8\%$ $/$ QP=22, $\Delta\!C_T\!\!\!=\!\!5\%$} & \multicolumn{2}{c|}{{QP=32, $\Delta\!C_T\!\!\!=\!\!18\%$ $/$ QP=22, $\Delta\!C_T\!\!\!=\!\!15\%$}} & \multicolumn{2}{c|}{QP=32, $\Delta\!C_T\!\!\!=\!\!23\%$ $/$ QP=22, $\Delta\!C_T\!\!\!=\!\!20\%$} \\
\cline{4-9}      &       &       & $\Delta\text{PSNR}$ & $\Delta\text{EW-PSNR}$ & $\Delta\text{PSNR}$ & $\Delta\text{EW-PSNR}$ & $\Delta\text{PSNR}$ & $\Delta\text{EW-PSNR}$ \\
\hline
\multirow{6}[4]{*}{A} & \multirow{3}[2]{*}{\textit{Traffic}} & SGCC  & 0.0848 / \textbf{0.0642}  & \textbf{0.0634} /\textbf{0.0478} & \textbf{0.2962} / \textbf{0.5657} & \textbf{0.2015} / \textbf{0.2863} & \textbf{1.0616} / \textbf{5.7466} & \textbf{0.5667} / \textbf{4.5571} \\
     &       & \cite{nogues2014power} & 0.3401 / 0.7851  & 0.4313 / 1.0425 & 1.0019 /2.2343 & 1.2724 /2.8351 & -\ \ \quad/\ \ \quad-  & -\ \ \quad/\ \ \quad- \\
      &       & \cite{nogues2015modified} & \textbf{0.0610} / 0.1213 & 0.0672 / 0.1507 & 0.9543 / 2.2104 & 1.2287 / 2.8138 & 9.3028 / 13.7660 & 10.3273 / 15.0217 \\
\cline{2-9}      & \multirow{3}[2]{*}{\textit{PeopleOnStreet}} & SGCC  & 0.1495 / \textbf{0.1236} & \textbf{0.0611} / \textbf{0.0502} & \textbf{0.4429} / \textbf{0.6011} & \textbf{0.4127} / \textbf{0.3580} & \textbf{0.8364} / \textbf{4.9234} & \textbf{0.6239} / \textbf{3.5759} \\
     &       & \cite{nogues2014power} & 0.3522 / 0.6946  & 0.3952 / 0.8403& 0.9160 / 1.7966  & 1.0259 / 2.1468  & -\ \ \quad/\ \ \quad-    & -\ \ \quad/\ \ \quad- \\
      &       & \cite{nogues2015modified} & \textbf{0.1135} / 0.1753 & 0.1232 / 0.1986 & 0.8036 / 1.7145 & 0.9098 / 2.0734 & 7.6765 / 12.4255 & 8.2545 / 13.4957 \\
\hline
\multirow{9}[6]{*}{B} & \multirow{3}[2]{*}{\textit{ParkScene}} & SGCC  & \textbf{0.0284} / \textbf{0.0451} & \textbf{0.0170} / \textbf{0.0612} & \textbf{0.4093} / \textbf{0.9248} & \textbf{0.3670} / \textbf{0.5026} & \textbf{1.0500} / \textbf{5.4168} & \textbf{0.7274} / \textbf{4.6220} \\
      &       & \cite{nogues2014power} & 0.2288 / 0.6820 & 0.2939 / 0.7777 & 0.5807 / 1.6772 & 0.6801 / 1.7581 & -\ \ \quad/\ \ \quad-    & -\ \ \quad/\ \ \quad- \\
     &       & \cite{nogues2015modified} & 0.0413 / 0.1047 & 0.0721 / 0.1349 & 0.5559 / 1.6712 & 0.6263 / 1.7338 & 6.5633 / 11.1630 & 6.5933 / 11.4560  \\
\cline{2-9}      & \multirow{3}[2]{*}{\textit{BQTerrace}} & SGCC  & \textbf{0.0152} / \textbf{0.0203} & \textbf{0.0069} / \textbf{0.0045} & \textbf{0.3236} / \textbf{1.1269} & \textbf{0.0601} / \textbf{0.3316} & \textbf{1.7257} / \textbf{6.9894} & \textbf{0.8471} / \textbf{6.1868} \\
     &       & \cite{nogues2014power} & 0.4375 / 0.9351 & 0.4674 / 1.1097 & 1.2279 / 2.4650 & 1.2851 / 2.8457 & -\ \ \quad/\ \ \quad-   & -\ \ \quad/\ \ \quad- \\
    &       & \cite{nogues2015modified} & 0.0599 / 0.1783 & 0.0604 / 0.1936 & 1.2235 / 2.4526 & 1.2826 / 2.8342  & 9.8686 / 13.1676 & 10.0341 / 13.9260 \\
\cline{2-9}      & \multirow{3}[2]{*}{\textit{Kimono}} & SGCC  & 0.1061 / \textbf{0.0846} & \textbf{0.0528} / \textbf{0.0604} & \textbf{0.3054} / \textbf{0.3357}& \textbf{0.3882} / \textbf{0.4299} & \textbf{0.4501} / \textbf{2.1454} & \textbf{0.5394} / \textbf{2.0825} \\
      &       & \cite{nogues2014power} & 0.2199 / 0.3527 & 0.2402 / 0.3845 & 0.5364 / 0.8965 & 0.5853 / 0.9659 & -\ \ \quad/\ \ \quad-    & -\ \ \quad/\ \ \quad- \\
     &       & \cite{nogues2015modified} & \textbf{0.0780} / 0.0901 & 0.1010 / 0.0993 & 0.4741 / 0.8571 & 0.5107 / 0.8993 & 4.5654 / 7.8190  & 4.5623 / 7.6625 \\
\hline
\multirow{6}[4]{*}{C} & \multirow{3}[2]{*}{\textit{RaceHorses}} & SGCC  & \textbf{0.0512} / \textbf{0.0909} & \textbf{0.0907} / \textbf{0.0816} & \textbf{0.3240} \ \textbf{0.9005} & \textbf{0.3779} / \textbf{0.8770} & \textbf{0.8482} / \textbf{4.0999} & \textbf{0.8542} / \textbf{3.6421} \\
     &       & \cite{nogues2014power} & 0.3050 / 0.9234 & 0.3316 / 0.8741  & 0.7213 / 2.0427 & 0.8007 / 1.9319  & -\ \ \quad/\ \ \quad-    & -\ \ \quad/\ \ \quad- \\
      &       & \cite{nogues2015modified} & 0.0858 / 0.1999 & 0.0938 / 0.2091 & 0.6687 / 2.0220 & 0.7387 / 1.9026 & 6.3171 / 11.1323 & 6.7362 / 11.3829 \\
\cline{2-9}      & \multirow{3}[2]{*}{\textit{PartyScene}} & SGCC  & \textbf{0.0168} / \textbf{0.0148} & \textbf{0.0076} / \textbf{0.0063} & \textbf{1.0892} / \textbf{4.1913} & \textbf{0.2351} / \textbf{0.7267} & \textbf{3.4195} / \textbf{9.5478} & \textbf{1.0783} / \textbf{6.4582} \\
     &       & \cite{nogues2014power} & 0.8385 / 2.7252 & 0.4663 / 1.7229 & 1.8108 / 5.2526 & 1.0914 / 3.6297 & -\ \ \quad/\ \ \quad-    & -\ \ \quad/\ \ \quad- \\
     &       & \cite{nogues2015modified} & 0.1147 / 0.4213 & 0.0762 / 0.2541  & 1.8182 / 5.2595 & 1.0728 / 3.6390 & 9.1921 / 15.2457 & 7.6641 / 13.9324 \\
\hline
\multirow{12}[6]{*}{D} & \multirow{3}[2]{*}{\textit{RaceHorses}} & SGCC  & \textbf{0.0783} / \textbf{0.0542} & \textbf{0.0634} / \textbf{0.0574} & \textbf{0.2969} / \textbf{1.6244} & \textbf{0.2432} / \textbf{0.7723} & \textbf{0.8350} / \textbf{6.7330} & \textbf{0.7986} / \textbf{6.5464} \\
     &       & \cite{nogues2014power} & 0.3312 / 1.1740  & 0.3912 / 1.4015 & 0.7623 / 2.5856 & 0.8941 / 2.9178 & -\ \ \quad/\ \ \quad-     & -\ \ \quad/\ \ \quad- \\
     &       & \cite{nogues2015modified} & 0.0785 / 0.1935 & 0.0831 / 0.2079  & 0.7108 / 2.5802 & 0.8303 / 2.8968 & 6.8314 / 13.1843  & 7.2170 / 13.5511 \\
\cline{2-9}      & \multirow{3}[2]{*}{\textit{BQSquare}} & SGCC  & \textbf{0.0010} / \textbf{0.0041} & \textbf{0.0052} / \textbf{0.0039} & \textbf{1.5612} / \textbf{6.6972} & \textbf{0.7684} / \textbf{4.0648} & \textbf{5.0606} / \textbf{13.8470} & \textbf{3.9165} / \textbf{13.0618} \\
      &       & \cite{nogues2014power} & 1.4420 / 3.6301 & 1.2487 / 3.3762 & 3.0622 / 6.8422 & 2.7089 / 6.4708  & -\ \ \quad/\ \ \quad-     & -\ \ \quad/\ \ \quad- \\
      &       & \cite{nogues2015modified} & 0.1907 / 0.5886 & 0.1603 / 0.5216 & 3.0669 / 6.8397 & 2.7118 / 6.4708 & 11.9616 / 17.6653 & 11.5873 / 17.5664\\
\cline{2-9}      & \multirow{3}[2]{*}{\textit{BlowingBubbles}} & SGCC  & \textbf{0.0181} / \textbf{0.0125} & \textbf{0.0101} / \textbf{0.0144} & \textbf{1.1739} / 5.3492 & \textbf{0.4200} / \textbf{2.7555} & \textbf{3.2883} / \textbf{11.2444} & \textbf{2.0604} / \textbf{10.0694} \\
      &       & \cite{nogues2014power} & 0.5719 / 2.0708 & 0.4786 / 2.0190 & 1.2774 / \textbf{4.1821} & 1.1231 / 4.1169 & -\ \ \quad/\ \ \quad-     & -\ \ \quad/\ \ \quad- \\
     &       & \cite{nogues2015modified} & 0.0639 / 0.2993 & 0.0527 / 0.2788 & 1.2869 / 4.1948 & 1.1302 / 4.1311 & 8.0493 / 14.1225 & 7.9817 / 14.4524 \\
\cline{2-9}      & \multirow{3}[2]{*}{\textit{BasketballPass}} & SGCC  & \textbf{0.0945} / \textbf{0.0567} & \textbf{0.0584} / \textbf{0.0332} & \textbf{0.3108} / \textbf{1.0316} & \textbf{0.2516} / \textbf{0.3296} & \textbf{0.8332} / \textbf{5.6245} & \textbf{0.5587} / \textbf{4.8829} \\
      &       & \cite{nogues2014power} & 0.3041 / 1.0168  & 0.3048 / 1.0379 & 0.7798 / 2.1075 & 0.7836 / 2.1202 & -\ \ \quad/\ \ \quad-     & -\ \ \quad/\ \ \quad- \\
     &       & \cite{nogues2015modified} & 0.0968 / 0.1751 & 0.0913 / 0.1934 & 0.7106 / 2.0762 & 0.7226 / 2.0924 & 5.5834 / 11.2566 & 5.3963 / 11.2870 \\
\hline
\multicolumn{2}{|c|}{\multirow{3}[2]{*}{\textbf{Average}}} & SGCC  & \textbf{0.0585} / \textbf{0.0519} & \textbf{0.0397} / \textbf{0.0383} & \textbf{0.5939} / \textbf{2.1226} & \textbf{0.3387} / \textbf{1.0395} & \textbf{1.7644} / \textbf{6.9380} & \textbf{1.1428} / \textbf{5.9714} \\
\multicolumn{2}{|c|}{} & \cite{nogues2014power} & 0.4883 / 1.3629 & 0.4590 / 1.3260 & 1.1524 / 2.9166 & 1.1137 / 2.8854  & -\ \ \quad/\ \ \quad-     & -\ \ \quad/\ \ \quad-  \\
\multicolumn{2}{|c|}{} & \cite{nogues2015modified} & 0.0895 / 0.2316 & 0.0892 / 0.2220  & 1.1158 / 2.8980 & 1.0695 / 2.8626  & 7.8101 / 12.8134 & 7.8504 / 13.0667 \\
\hline
\end{tabular}%
\vspace{-2em}
\end{table*}

\vspace{-1em}
\subsection{Evaluation on control performance}

First of all, we evaluate the control performance of our SGCC approach in HEVC decoding. The performance evaluation consists of two parts: Maximal Achievable Reduction (MAR) and control error. First, we compare the MAR results of our SGCC approach with those of \cite{nogues2014power} and \cite{nogues2015modified}. Here, to obtain MAR of our SGCC approach, we set $f_n = 1$ and $g_n = 3$ for all the CTUs to achieve the maximal decoding complexity reduction. Then, we record the ratio of such reduction as MAR. For \cite{nogues2014power} and \cite{nogues2015modified}, we also make their complexity reduction reach maximal values using the ways reported in \cite{nogues2014power} and \cite{nogues2015modified}. Note that the complexity overhead of our approach (analyzed in Section \ref{overhead}), which is far less than HEVC decoding complexity, is included for evaluation.

\textbf{MAR:} Fig. \ref{fig:reduction} demonstrates the mean and standard deviation of MARs for our and conventional approaches. Here, the mean and standard deviation are calculated over all 11 test sequences at QP = 22, 27, 32 and 37. We can see from Fig. \ref{fig:reduction} that the MARs of our approach are much larger than those of \cite{nogues2014power} and \cite{nogues2015modified}. Specifically, the averaged MARs of our approach are $24.9\%$, $32.2\%$, $36.0\%$ and $39.5\%$, corresponding to QP = 22, 27, 32 and 37. By contrast, the averaged MARs of \cite{nogues2015modified} only reach $18.4\%$, $21.2\%$, $23.0\%$ and $23.4\%$ for QP = 22, 27, 32 and 37. Unfortunately, \cite{nogues2014power} obtains even less MARs. We can also see from Fig. \ref{fig:reduction} that larger MAR can be achieved in our SGCC approach, alongside increased QP. It is mainly due to the fact that small QP leads to much more coding bits, making entropy decoding consume higher complexity. However, even in the worst case of QP = 22, our approach has $24.9\%$ MAR in average, whereas the MARs of \cite{nogues2014power} and \cite{nogues2015modified} are $14.4\%$ and $18.4\%$, respectively.

Table \ref{tab:contr} reports the individual contributions of DF disabling and MC simplification in terms of MAR. It can be seen that disabling DF of all CTUs can averagely reduce HEVC decoding complexity by $12.5\%$, $15.3\%$, $16.0\%$ and $14.3\%$, corresponding to QP = 22, 27, 32 and 37, respectively. MC simplification is able to further achieve $12.4\%$, $16.9\%$, $20.0\%$ and $25.2\%$ complexity reduction. As such, the averaged MAR of our SGCC approach is able to reach $24.9\%$, $32.2\%$, $36.0\%$ and $39.5\%$, respectively, by both disabling DF and simplifying MC.

\textbf{Control error:} Next, we move to the evaluation of control error for our SGCC approach. Note that we do not compare with \cite{nogues2014power} and \cite{nogues2015modified} in control error, since \cite{nogues2014power} and \cite{nogues2015modified} are complexity reduction approaches, rather than complexity control. Table \ref{tab:complexity} reports the control errors of each sequence across different complexity reduction targets (i.e., $\Delta C_T = 10\%, 20\%, 30\%\ \text{and}\ 40\%$), at QP = 22, 27, 32 and 37. We can see from this table that in our approach the control error is up to $7.30\%$, while most errors are below $4.00\%$. Table \ref{tab:complexity} also tabulates Mean Absolute Error (MAE) and Mean Relative Error (MRE) for each specific $\Delta C_T$, averaged over all 11 test sequences.
It is apparent that MAEs of our approach in almost all cases are below $3.00\%$. The only exception is MAE = $3.42\%$, when $\Delta C_T$ is as large as $40\%$. Indeed, it is also necessary to evaluate MRE at different $\Delta C_T$, calculated by
\begin{equation}
\small
\text{MRE} = \frac{\text{MAE}}{\Delta C_T} \times 100\%,
\end{equation}
which indicates the proportion of control error with respect to $\Delta C_T$.
We can further see from Table \ref{tab:complexity} that MREs of most cases are less than $10\%$. In a conclusion, our SGCC approach performs well in control accuracy.

Note that when $\Delta C_T$ is less than the MAR of DF disabling (as shown in Table \ref{tab:contr}), HEVC decoding complexity is reduced by only disabling DF of some CTUs. When $\Delta C_T$ cannot be reached by disabling DF of all the CTUs, the remaining amount of decoding complexity reduction is contributed by MC simplification.

\vspace{-1em}
\subsection{Evaluation on complexity-distortion performance}
Now, we compare complexity-distortion performance of our SGCC approach with conventional approaches \cite{nogues2014power} and \cite{nogues2015modified}. The quality loss caused by decoding complexity reduction is measured in terms of $\Delta \text{PSNR}$ and $\Delta \text{EW-PSNR}$. $\Delta \text{PSNR}$ reflects objective quality loss, while $\Delta \text{EW-PSNR}$ measures perceptual quality loss. Table \ref{tab:psnr&w-psnr} shows $\Delta \text{PSNR}$ and $\Delta \text{EW-PSNR}$ of our and other conventional approaches, when $\Delta C_T = 8\%, 18\%\ \text{and}\ 23\%$\footnote{Since \cite{nogues2014power} and \cite{nogues2015modified} cannot control decoding complexity reduction, we were not able to set complexity reduction target $\Delta C_T$ in \cite{nogues2014power} and \cite{nogues2015modified}. Instead, we first decoded the test sequences with \cite{nogues2014power} and \cite{nogues2015modified}, and we found that their complexity reduction is around some specific values, e.g., 5\%, 10\% and 20\% at QP = 22, and 8\%, 18\% and 23\% at QP = 32. Then, we set $\Delta C_T$ of our SGCC approach to these values for fair comparison.}. Due to space limitation, the results of QP = 22 and 32 are provided in Table \ref{tab:psnr&w-psnr}.

\begin{figure*}[!t]
\centering
\vspace{-.5em}
  \subfigure{\includegraphics[width=.5\linewidth]{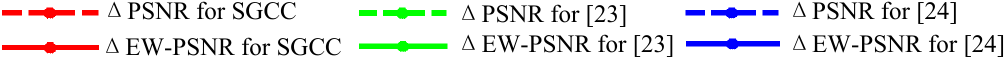}}\\
  \vspace{-.5em}
 \subfigure{\includegraphics[width=.9\linewidth]{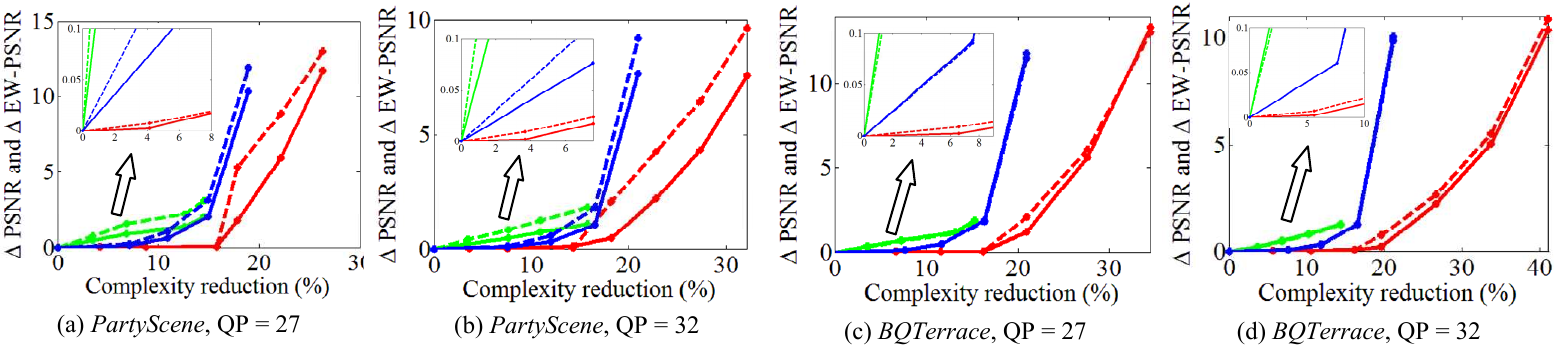}}
\vspace{-.5em}
\caption{\footnotesize{$\Delta\text{PSNR}$ and $\Delta\text{EW-PSNR}$ versus decoding complexity reduction at QP = 27 and 32.}}\label{fig:ew_psnr}
\end{figure*}

\textbf{Objective quality loss:} It can be seen from Table \ref{tab:psnr&w-psnr} that Y-PSNR loss of our SGCC and other approaches increase dramatically, when decoding complexity reduction becomes larger. For example, when decoding complexity reduction increases from 8\% to 18\% at QP = 32, the averaged $\Delta \text{PSNR}$ of our SGCC approach enhances from 0.0585dB to 0.5939dB. Once complexity reduction reaches 23\%, $\Delta \text{PSNR}$ of our approach increases to 1.7644dB. It is because MC simplification of (22-b) brings in larger distortion, in comparison with DF disabling of (22-a). It can be further seen from Table \ref{tab:psnr&w-psnr} that our SGCC approach significantly outperforms \cite{nogues2014power} and \cite{nogues2015modified} in terms of $\Delta \text{PSNR}$, especially at high complexity reduction. Specifically, once decoding complexity reduction increases to 23\%, \cite{nogues2015modified} incurs averagely 7.8504dB Y-PSNR loss at QP = 32, far more than 1.7644dB of our SGCC approach. Besides, \cite{nogues2014power} is incapable of reducing decoding complexity of HEVC to 23\%. Despite much better than  \cite{nogues2014power} and \cite{nogues2015modified}, the objective quality loss of our method is not very small at high complexity reduction (e.g., $\Delta \text{PSNR}=1.7644$ dB at 23\% reduction and QP = 32). However, the perceptual quality loss by our method can be alleviated (e.g., $\Delta \text{EW-PSNR}=1.1428$ dB at $23 \%$ reduction and QP = 32), which is the minimization objective of our SGCC approach. For QP = 22, similar results can be found from Table \ref{tab:psnr&w-psnr}.

\textbf{Perceptual quality loss:} Table \ref{tab:psnr&w-psnr} shows that, for all 11 test sequences across three decoding complexity targets, $\Delta \text{EW-PSNR}$ of our SGCC approach is less than those of \cite{nogues2014power} and \cite{nogues2015modified}. For example, when $\Delta C_T = 18\%$ and QP = 32, averaged $\Delta \text{EW-PSNR}$ is 0.3387 dB, 1.1137 dB and 1.0695 dB for the SGCC, \cite{nogues2014power} and \cite{nogues2015modified} approaches. This implies better perceptual quality achieved by our SGCC approach. Furthermore, the averaged values of $\Delta \text{EW-PSNR}$ are much less than those of $\Delta \text{PSNR}$ in our SGCC approach, while in \cite{nogues2014power} and \cite{nogues2015modified} the values of $\Delta \text{EW-PSNR}$ are similar to those of $\Delta \text{PSNR}$. For example, the averaged $\Delta \text{PSNR}$ of our approach is 0.5939dB at $\Delta C_T = 18\%$ and QP = 32, while $\Delta \text{EW-PSNR}$ is averagely 0.3387 dB. In contrast, the averaged values of $\Delta \text{EW-PSNR}$ and $\Delta \text{PSNR}$ are 1.1524 dB and 1.1137dB for \cite{nogues2014power}, and 1.1158 dB and 1.0695 dB for \cite{nogues2015modified}, at $\Delta C_T = 18\%$ and QP = 32. As shown in Table \ref{tab:psnr&w-psnr}, our SGCC approach also performs well in perceptual quality at QP = 22. In a word, the above results verify that our approach is capable of optimizing perceptual quality, when the decoding complexity of HEVC is reduced.

\textbf{Complexity-reduction curves:} To investigate the quality loss at varying reduction of decoding complexity, Fig. \ref{fig:ew_psnr} plots the complexity-distortion curves of five selected test sequences, for our SGCC and other conventional approaches. We provide in this figure the complexity-distortion curves of QP = 27 and 32 to show the generalization of our approach at different bit rates. In Fig. \ref{fig:ew_psnr}, the curves for both $\Delta \text{PSNR}$ and $\Delta \text{EW-PSNR}$ are shown, which reflect the objective and perceptual quality loss, respectively. As shown in this figure, both $\Delta \text{PSNR}$ and $\Delta \text{EW-PSNR}$ of our SGCC approach are less than those of \cite{nogues2014power} and \cite{nogues2015modified}. Besides, we can observe that $\Delta \text{EW-PSNR}$ is less than $\Delta \text{PSNR}$ in our SGCC approach, indicating better perceptual quality.

\begin{figure}
\centering
  \subfigure{\includegraphics[width=1\linewidth]{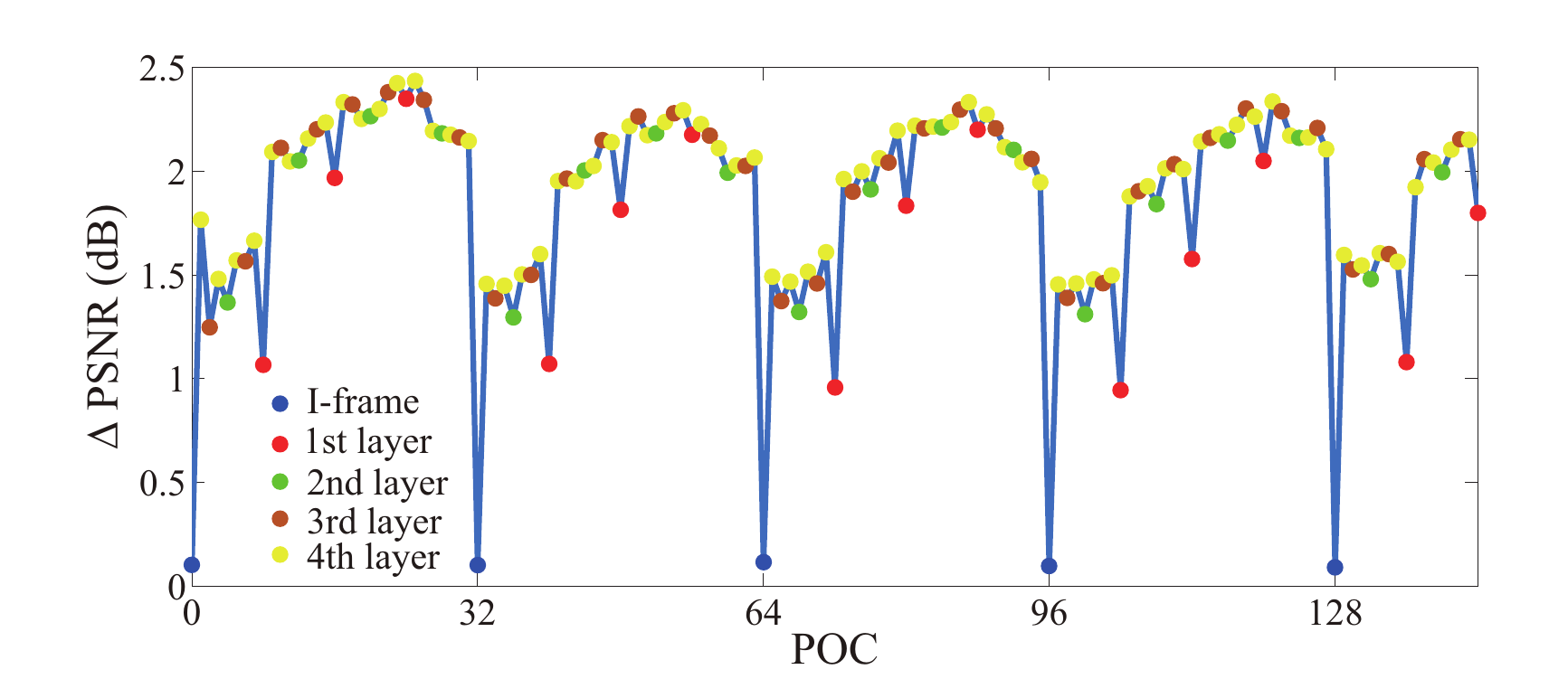}}\\
  \vspace{-1.3em}
  \hspace{.17em} \subfigure{\includegraphics[width=1\linewidth]{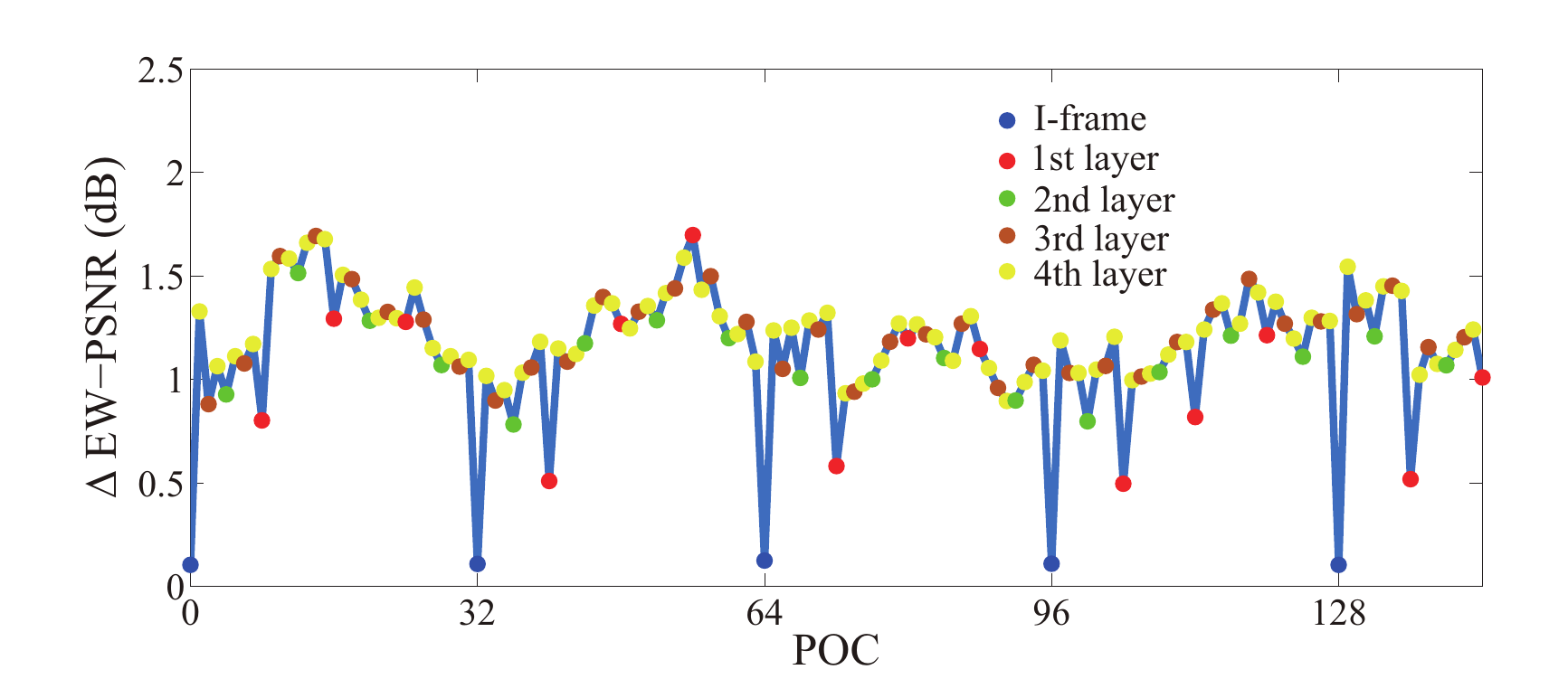}}
  \vspace{-1em}
\caption{\footnotesize{The frame-level $\Delta \text{PSNR}$ and $\Delta \text{EW-PSNR}$ when $\Delta C_T = 23\%$ and QP = 32. }}\label{fig:f_psnr2}
\vspace{-1.5em}
\end{figure}

\begin{table*}[!t]
\vspace{-1em}
\caption{\footnotesize{DMOS values at QP = 32 of SGCC, \cite{nogues2014power} and \cite{nogues2015modified}.}} \label{tab:DMOS}
\centering
\scriptsize
\begin{tabular}{|c|c|r|r|r|r|r|r|r|r|r|r|r|c|}
\hline
$\Delta C_T$ & Sequences & \multicolumn{1}{c|}{1} & \multicolumn{1}{c|}{2} & \multicolumn{1}{c|}{3} & \multicolumn{1}{c|}{4} & \multicolumn{1}{c|}{5} & \multicolumn{1}{c|}{6} & \multicolumn{1}{c|}{7} & \multicolumn{1}{c|}{8} & \multicolumn{1}{c|}{9} & \multicolumn{1}{c|}{10} & \multicolumn{1}{c|}{11} & \textbf{Average} \\
\hline
\multirow{3}[2]{*}{$8\%$} & SGCC  & \textbf{33.35 } & 43.75  & \textbf{42.41 } & \multicolumn{1}{c|}{\textbf{37.50 }} & \textbf{45.03 } & \multicolumn{1}{c|}{53.51 } & \textbf{41.30 } & \multicolumn{1}{c|}{52.06 } & \multicolumn{1}{c|}{\textbf{33.20 }} & \multicolumn{1}{c|}{\textbf{37.07 }} & \textbf{33.05 } & \textbf{41.11 } \\
\cline{2-14}      & [23]  & 40.79  & 44.28  & 44.25  & 43.47  & 45.79  & \multicolumn{1}{c|}{51.40 } & 46.20  & \multicolumn{1}{c|}{49.97 } & \multicolumn{1}{c|}{37.90 } & \multicolumn{1}{c|}{41.76 } & 44.70  & 44.59  \\
\cline{2-14}      & [24]  & 47.26  & \textbf{36.65 } & 44.68  & 45.27  & 47.69  & \multicolumn{1}{c|}{\textbf{47.97 }} & 46.88  & \multicolumn{1}{c|}{\textbf{46.07 }} & \multicolumn{1}{c|}{35.07 } & \multicolumn{1}{c|}{42.05 } & 41.55  & 43.74  \\
\hline
\multirow{2}[4]{*}{$23\%$} & SGCC  & \textbf{45.98 } & \textbf{47.23 } & \textbf{61.06 } & \textbf{43.33 } & \textbf{54.81 } & \textbf{54.43 } & \textbf{55.07 } & \textbf{60.83 } & \textbf{63.06 } & \textbf{64.37 } & \textbf{48.22 } & \textbf{54.40 } \\
\cline{2-14}      & [24]  & 65.71  & 65.07  & 58.44  & 70.11  & 57.50  & 66.11  & 69.80  & 70.87  & 68.08  & 72.93  & 63.09  & 66.16  \\
\hline
\multicolumn{14}{c}{\scriptsize{1: \emph{Traffic} 2: \emph{PeopleOnStreet} 3: \emph{ParkScene} 4: \emph{BQTerrace} 5: \emph{Kimono} 6: \emph{RaceHorses} $(832\times 480)$ }}\\
\multicolumn{14}{c}{\scriptsize{7: \emph{PartyScene} 8: \emph{RaceHorses} $(416\times 240)$ 9: \emph{BQSquare} 10: \emph{BlowingBubbles} 11: \emph{BasketballPass}}}\\
\end{tabular}%
\end{table*}

\begin{figure*}[!t]
\vspace{-1em}
\centering
  \includegraphics[width=.9\linewidth]{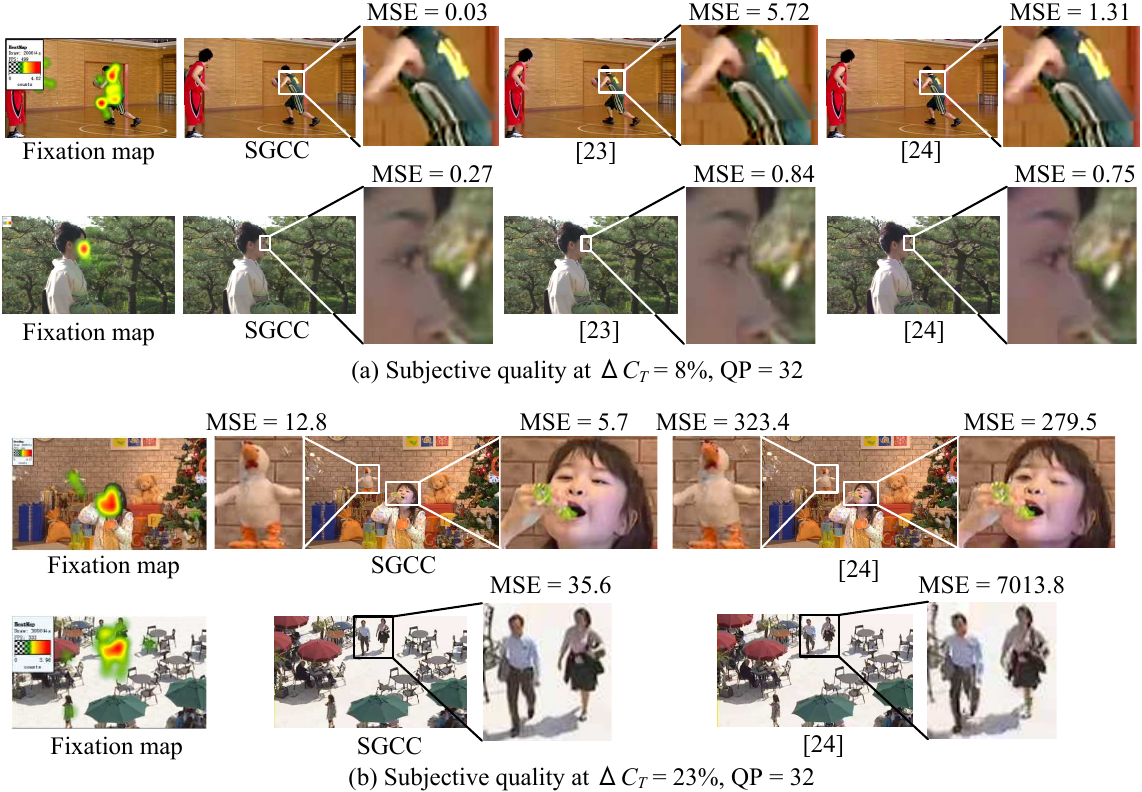}
\vspace{-1em}
\caption{\footnotesize{Subjective quality of four selected frames decoded by HEVC with our SGCC, \cite{nogues2014power} and \cite{nogues2015modified} approaches, at $\Delta C_T = 8\% $ and $\Delta C_T = 23\% $. The MSEs of ROI in the four selected frames are given. The MSEs of our SGCC approach are significantly smaller than those of \cite{nogues2014power} and \cite{nogues2015modified}. }}\label{fig:zhuguantu}
\vspace{-1em}
\end{figure*}

\subsection{Assessment on fluctuation of quality loss}
Next, we assess the frame-level fluctuation of quality loss caused by our SGCC approach, since the error propagation of our approach may increase the fluctuation of quality loss. Fig. \ref{fig:f_psnr2} plots the objective and perceptual quality loss along with decoded frames at $\Delta C_T = 23\%$ and QP = 32, averaged over all 11 test sequences. First, it can be seen that I-frames have slight quality loss, which incur no error propagation.
More importantly, the quality loss can be resumed to be near zero successively after I frames, validating the effectiveness of I frames in preventing error propagation of quality loss. This is in accordance with Observation \ref{ob1}. Second, the quality degradation of the frames at the first layer is less than that at upper layers, within a GOP. As such, the fluctuation of quality loss can be relieved. This indicates the small error propagation of our approach due to the hierarchical coding structure of HEVC, satisfying Observation \ref{ob2}. Finally, one may see that the range of $\Delta\text{EW-PSNR}$ (0.5-1.7 dB) is much smaller than that of $\Delta \text{PSNR}$ (1-2.5 dB), for non-I frames. Thus, it verifies that the perceptual quality loss of our SGCC approach has less fluctuation, compared with objective quality loss.

\vspace{-1em}
\subsection{Assessment on subjective quality}

\begin{table*}[!t]
  \centering
  \scriptsize
  \vspace{-1em}
  \caption{\footnotesize{Complexity control error (\%) of our SGCC approach for HEVC bitstreams with rate control.}}
    \begin{tabular}{|c|c|c|c|c|c|c|c|c|c|c|c|c|}
    \hline
    \multirow{3}[6]{*}{\textbf{Classes}} & \multirow{3}[6]{*}{\textbf{Sequences}} & \multicolumn{2}{c|}{Bit rate 1} & \multicolumn{2}{c|}{Bit rate 2} & \multicolumn{3}{c|}{Bit rate 3} & \multicolumn{4}{c|}{Bit rate 4} \\
\cline{3-13}          &       & \multicolumn{2}{c|}{$\Delta C_T$ (\%)} & \multicolumn{2}{c|}{$\Delta C_T$ (\%)} & \multicolumn{3}{c|}{$\Delta C_T$ (\%)} & \multicolumn{4}{c|}{$\Delta C_T$ (\%)} \\
\cline{3-13}          &       & \textbf{10} & \textbf{20} & \textbf{10} & \textbf{20} & \textbf{10} & \textbf{20} & \textbf{30} & \textbf{10} & \textbf{20} & \textbf{30} & \textbf{40} \\
    \hline
    \multirow{2}[4]{*}{A} & \textit{Traffic} & +1.02  & +1.08  & +0.45  & -0.28  & +0.43  & -0.82  & -0.25  & +0.62  & -1.45  & +0.79  & +2.63  \\
\cline{2-13}          & \textit{PeopleOnStreet} & +1.79  & -0.96  & +2.56  & +1.67  & +2.57  & +3.79  & -1.20  & +1.77  & +3.86  & +0.76  & -2.50  \\
    \hline
    \multirow{3}[6]{*}{B} & \textit{ParkScene} & +0.90  & -0.34  & +1.11  & -0.67  & +1.50  & -0.62  & -0.80  & +0.71  & -0.83  & +0.26  & +1.53  \\
\cline{2-13}          & \textit{BQTerrace} & -2.45  & -4.39  & -0.03  & -1.78  & +0.88  & -1.75  & -0.18  & +0.76  & -2.50  & +0.40  & +2.79  \\
\cline{2-13}          & \textit{Kimono} & +0.82  & -0.95  & +0.63  & -1.12  & +0.08  & -1.12  & -1.29  & -0.09  & -1.74  & -0.70  & -0.01  \\
    \hline
    \multirow{2}[4]{*}{C} & \textit{RaceHorses} & -0.63  & -4.89  & +1.34  & -1.79  & +3.03  & +0.47  & -5.49  & +2.58  & +1.26  & -2.76  & -6.91  \\
\cline{2-13}          & \textit{PartyScene} & -2.45  & -5.32  & -1.44  & -3.48  & -0.04  & -2.22  & -4.96  & -0.25  & -2.17  & -2.60  & -3.40  \\
    \hline
    \multirow{4}[8]{*}{D} & \textit{RaceHorses} & -0.97  & -4.18  & -0.08  & -1.15  & +0.12  & +0.04  & -5.24  & +1.72  & +1.12  & -2.08  & -6.73  \\
\cline{2-13}          & \textit{BQSquare} & -3.16  & -4.01  & -2.17  & -3.48  & -2.42  & -3.72  & -3.56  & -1.96  & -3.70  & -1.71  & -1.17  \\
\cline{2-13}          & \textit{BlowingBubbles} & -3.88  & -5.00  & -2.69  & -3.62  & -2.28  & -3.65  & -5.29  & -1.87  & -2.98  & -2.91  & -4.50  \\
\cline{2-13}          & \textit{BasketballPass} & -1.71  & -3.93  & -1.00  & -1.92  & -0.58  & -0.52  & -4.32  & -0.24  & 0.06  & -2.16  & -5.69  \\
    \hline
    \multicolumn{2}{|c|}{\textbf{MAE}} & \textbf{1.80} & \textbf{3.19} & \textbf{1.23} & \textbf{1.91} & \textbf{1.27} & \textbf{1.70} & \textbf{2.96} & \textbf{1.14} & \textbf{1.97} & \textbf{1.56} & \textbf{3.44} \\
    \hline
    \multicolumn{2}{|c|}{\textbf{MRE}} & \textbf{17.99} & \textbf{15.93} & \textbf{12.28} & \textbf{9.53} & \textbf{12.68} & \textbf{8.51} & \textbf{9.88} & \textbf{11.41} & \textbf{9.85} & \textbf{5.19} & \textbf{8.61} \\
    \hline
    \end{tabular}%
  \label{tab:rc}%
\end{table*}%

\begin{table*}[!t]
\scriptsize
\centering
\hspace{-2.5em}
\caption{\footnotesize{$\Delta \text{PSNR}$ and $\Delta \text{EW-PSNR}$ (dB) of our SGCC approach for HEVC bitstreams with rate control.}}
\begin{tabular}{|c|c|c|c|c|c|c|c|}
\hline
\multirow{2}[2]{*}{ {Class}} & \multirow{2}[2]{*}{ {Sequence}} & \multicolumn{2}{c|}{Bitrate3, $\Delta\!C_T\!\!\!=\!\!8\%$$/$Bitrate1, $\Delta\!C_T\!\!\!=\!\!5\%$} & \multicolumn{2}{c|}{{Bitrate3, $\Delta\!C_T\!\!\!=\!\!18\%$$/$Bitrate1, $\Delta\!C_T\!\!\!=\!\!15\%$}} & \multicolumn{2}{c|}{Bitrate3, $\Delta\!C_T\!\!\!=\!\!23\%$$/$Bitrate1, $\Delta\!C_T\!\!\!=\!\!20\%$} \\
\cline{3-8}      &             & $\Delta\text{PSNR}$ & $\Delta\text{EW-PSNR}$ & $\Delta\text{PSNR}$ & $\Delta\text{EW-PSNR}$ & $\Delta\text{PSNR}$ & $\Delta\text{EW-PSNR}$ \\
\hline
\multirow{2}[4]{*}{A} & {\textit{Traffic}}    & 0.0618 / 0.0554 & 0.0418 / 0.0339 & 0.2949 / 0.6880 & 0.2012 / 0.3102 & 1.0134 / 5.4027 & 0.4770 / 3.8365\\
\cline{2-8}      & {\textit{PeopleOnStreet}}  & 0.1713 / 0.1141 & 0.0757 / 0.0460 & 0.5142 / 0.6207 & 0.4547 / 0.3026 & 0.8323 / 4.2424 & 0.6084 / 2.8400  \\
\hline
\multirow{3}[6]{*}{B} & {\textit{ParkScene}}   &  0.0582 / 0.0380 & 0.0762 / 0.0490 & 0.3926 / 1.0378 & 0.3504 / 0.6955 & 0.9385 / 4.8877 & 0.7082 / 3.9699  \\
\cline{2-8}      & {\textit{BQTerrace}}  & 0.0172 / 0.0175 & 0.0105 / 0.0086 & 0.4438 / 1.1507 & 0.1167 / 0.2280 & 1.8588 / 6.6625 & 0.8665 / 6.0677  \\
\cline{2-8}      & {\textit{Kimono}}   &  0.1595 / 0.1485 & 0.1121 / 0.0742 & 0.3395 / 0.3486 & 0.4164 / 0.3090 & 0.5632 / 1.9507 & 0.6139 / 1.7813  \\
\hline
\multirow{2}[4]{*}{C} & {\textit{RaceHorses}}   & 0.1077 / 0.1269 & 0.1081 / 0.1549 & 0.3323 / 0.7735 & 0.3195 / 0.4081 & 0.8656 / 5.2604 & 0.7179 / 4.5446   \\
\cline{2-8}      & {\textit{PartyScene}}   & 0.0310 / 0.0364 & 0.0233 / 0.0695 & 1.3267 / 3.8102 & 0.2805 / 0.9081 & 3.5181 / 10.7087 & 1.1746 / 6.7472  \\
\hline
\multirow{4}[6]{*}{D} & {\textit{RaceHorses}}   & 0.0909 / 0.0561 & 0.0706 / 0.0540 & 0.3112 / 1.2878 & 0.2393 / 0.4535 & 0.8652 / 5.6916 & 0.8331 / 5.2178  \\
\cline{2-8}      & {\textit{BQSquare}}  & 0.0206 / 0.0174 & 0.0147 / 0.0178 & 1.1670 / 4.7376 & 0.4542 / 3.3934 & 3.2781 / 10.6592 & 2.1648 / 10.3724   \\
\cline{2-8}      & {\textit{BlowingBubbles}}   & 0.0004 / 0.0036 & 0.0016 / 0.0036 & 1.4757 / 6.0610 & 0.8510 / 4.2087 & 4.7418 / 13.1899 & 3.7271 / 12.3745 \\
\cline{2-8}      & {\textit{BasketballPass}}   & 0.1109 / 0.0539 & 0.0582 / 0.0391 & 0.4368 / 0.9602 & 0.3012 / 0.4422 & 1.0124 / 4.8030 & 0.6371 / 4.1436  \\
\hline
\multicolumn{2}{|c|}{{ {Average}}}  & 0.0754 / 0.0607 & 0.0539 / 0.0500 & 0.6395 / 1.9524 & 0.3623 / 1.0599 & 1.7716 / 6.6781 & 1.1390 / 5.6268
\\
\hline
\end{tabular}\label{tab:ratepsnr}
\end{table*}

We further assess the subjective quality of our SGCC approach compared with \cite{nogues2014power} and \cite{nogues2015modified}. In our experiment, the DMOS test was conducted to rate subjective quality of the decoded sequences, by the means of Single Stimulus Continuous Quality Evaluation (SSCQE), which is processed by Rec. ITU-R BT.500 \cite{recommendation2002500}. During the test\footnote{Here, a Sony BRAVIA XDVW600 television, with a 55-inch LCD displaying screen, was utilized to display the decoded sequences. The viewing distance was set to be approximately four times of the video height for rational evaluation. The rating score includes excellent (100-81), good (80-61), fair (60-41), poor (40-21), and bad (20-1).}, sequences were displayed in random order. After viewing each decoded sequence, the subjects were asked to rate the sequence. As a result, DMOS value of each decoded sequence can be calculated to measure the difference of subjective quality between sequences decoded by original HEVC and by HEVC with our SGCC approach or other conventional approaches \cite{nogues2014power} and \cite{nogues2015modified}.

Table \ref{tab:DMOS} shows the DMOS values of three approaches for all test sequences, with complexity reduction being approximately 8\% and 23\%. Note that the smaller values of DMOS mean the better subjective quality, since DMOS quantifies the subjective quality difference between the uncompressed and compressed sequences. Obviously, when complexity reduction is around 8\%, our SGCC approach has smaller DMOS values than \cite{nogues2014power} and \cite{nogues2015modified} for 8 among 11 test sequences. Besides, the averaged DMOS value of our SGCC approach is smallest among all three approaches at $\Delta C_T = 8\%$. Once decoding complexity is further deceased to 23\%, our SGCC approach is greatly superior to \cite{nogues2015modified} for all 11 test sequences, in terms of DMOS. Recall that decoding complexity reduction of \cite{nogues2014power} cannot arrive at 23\%, and we thus only compare with \cite{nogues2015modified} for $\Delta C_T = 23\%$ in Table \ref{tab:DMOS}.

Furthermore, Fig. \ref{fig:zhuguantu} shows some frames of four selected sequences, decoded by HEVC with the SGCC, \cite{nogues2014power} and \cite{nogues2015modified} approaches. We can observe that the sequences by \cite{nogues2014power} and \cite{nogues2015modified} have severe blur and blocky artifacts in ROI, at $\Delta C_T = 8\%$. On the contrary, our SGCC approach results in better subjective quality with less blur and blocky artifacts. When $\Delta C_T$ is 23\%, our SGCC approach enjoys more obvious quality improvement over \cite{nogues2014power} and \cite{nogues2015modified}, as seen in Fig. \ref{fig:zhuguantu}-(b). This is in accord with the DMOS results above.

\subsection{Performance on HEVC bitstreams with other configurations}
In practical applications, the HEVC encoder usually enables rate control, so that frame-level QP values may vary within a sequence. Here, we further tested the trained parameters of our SGCC approach on the HEVC bitstreams encoded with rate control enabling. Here, parameters $a$, $b$ and $c$ are chosen according to the range of frame-level QP, as discussed in Section \ref{modelMC}. The results are shown in Tables \ref{tab:rc} and \ref{tab:ratepsnr}. It can be seen that the control accuracy of decoding complexity and the degradation of quality are comparable to those without rate control (Tables \ref{tab:complexity} and \ref{tab:psnr&w-psnr}). Note that we follow the most recent rate control work of \cite{li2014domain} to set the target bit rates the same as the actual bit rates at fixed QPs (22, 27, 32 and 37), as reported in Tables \ref{tab:rc} and \ref{tab:ratepsnr}.

We further tested the trained parameters on the HEVC bitstreams encoded with different GOP size. The results Tables \ref{tab:gop} and \ref{tab:goppsnr} show the results for GOP size of 4. As shown in these tables, the performance of control accuracy and quality loss of sequences are comparable to those with GOP size of 8 (Tables \ref{tab:complexity} and \ref{tab:psnr&w-psnr}). Therefore, the parameters in Table \ref{tab:coc} are effective for the HEVC bitstreams encoded with different GOP sizes. To summarize, the trained parameters of our SGCC approach are also applicable for different encoding configurations.

\begin{table*}[!t]
  \centering
  \scriptsize
  \vspace{-1em}
  \caption{\footnotesize{Complexity control error (\%) of our SGCC approach for HEVC bitstreams with GOP size as 4.}}
    \begin{tabular}{|c|c|c|c|c|c|c|c|c|c|c|c|c|c|}
    \hline
    \multirow{3}[6]{*}{{Classes}} & \multirow{3}[6]{*}{{Sequences}} & \multicolumn{2}{c|}{QP = 22} & \multicolumn{3}{c|}{QP = 27} & \multicolumn{3}{c|}{QP = 32} & \multicolumn{4}{c|}{QP = 37} \\
\cline{3-14}          &       & \multicolumn{2}{c|}{$\Delta C_T$ (\%)} & \multicolumn{3}{c|}{$\Delta C_T$ (\%)} & \multicolumn{3}{c|}{$\Delta C_T$ (\%)} & \multicolumn{4}{c|}{$\Delta C_T$ (\%)} \\
\cline{3-14}          &       & \textbf{10} & \textbf{20} & \textbf{10} & \textbf{20} & \textbf{30}&\textbf{10} & \textbf{20} & \textbf{30} & \textbf{10} & \textbf{20} & \textbf{30} & \textbf{40} \\
    \hline
    \multirow{2}[4]{*}{A} & \textit{Traffic} &-0.22  & -0.74  & -0.36  & +0.04  & +1.03  & +1.08  & -0.44  & -0.93  & +2.00  & +0.06  & +0.59  & +0.21 \\
\cline{2-14}          & \textit{PeopleOnStreet} &+1.76  & -2.19  & -0.60  & +1.54  & -2.28  & +0.58  & +1.48  & +0.39  & +0.62  & +3.47  & -0.45  & -4.94\\
    \hline
    \multirow{3}[6]{*}{B} & \textit{ParkScene} & +2.61  & +0.42  & -0.37  & -1.23  & +0.19  & +1.25  & -0.90  & +0.24  & +1.35  & -0.49  & +0.17  & -0.11  \\
\cline{2-14}          & \textit{BQTerrace} & -2.15  & -4.32  & +1.43  & +0.42  & +1.57  & +0.46  & +0.41  & -0.80  & +0.80  & +0.50  & +1.26  & +1.52\\
\cline{2-14}          & \textit{Kimono} & +1.93  & -0.41  & -1.42  & -0.65  & -1.84  & -0.72  & -0.51  & -2.65  & -1.20  & -0.30  & -2.96  & -3.21  \\
    \hline
    \multirow{2}[4]{*}{C} & \textit{RaceHorses} &-0.55  & -2.53  & +0.48  & -2.71  & -9.09  & -0.40  & -0.11  & -3.88  & +0.41  & -0.87  & -1.74  & -8.77 \\
\cline{2-14}          & \textit{PartyScene} & -1.49  & -2.08  & -0.64  & +0.25  & -4.38  & +1.12  & -1.21  & -2.79  & +1.68  & +0.37  & -1.04  & -5.87   \\
    \hline
    \multirow{4}[8]{*}{D} & \textit{RaceHorses} &-1.12  & -2.10  & -0.23  & -2.53  & -9.62  & +1.61  & +0.99  & -3.87  & -0.41  & 0.90  & -0.94  & -8.91  \\
\cline{2-14}          & \textit{BQSquare} & +1.37  & -0.10  & -1.51  & -0.89  & -3.14  & +0.05  & +0.08  & +0.49  & +1.13  & +0.65  & +1.15  & -2.35  \\
\cline{2-14}          & \textit{BlowingBubbles} & +1.07  & -1.81  & -1.23  & -2.60  & -7.56  & -0.06  & -0.83  & -3.15  & +0.85  & +0.60  & -1.34  & -7.40 \\
\cline{2-14}          & \textit{BasketballPass}& -0.75  & -1.18  & -0.75  & -2.02  & -8.34  & 0.67  & 0.53  & -3.30  & 1.43  & -0.21  & -1.32  & -8.68  \\
    \hline
    \multicolumn{2}{|c|}{\textbf{MAE}} & \textbf{1.25}  & \textbf{1.49}  &  \textbf{0.75} & \textbf{1.24} &\textbf{4.09}  & \textbf{0.67}  & \textbf{0.62}  &  \textbf{1.87} &  \textbf{0.99} & \textbf{0.70}  & \textbf{1.08}  & \textbf{4.33}\\
    \hline
    \multicolumn{2}{|c|}{\textbf{MRE}} & \textbf{12.52}  & \textbf{7.45}  &\textbf{7.51}   & \textbf{6.20}  & \textbf{13.62} & \textbf{6.67}  & \textbf{3.12}  &  \textbf{6.24} & \textbf{9.90}  & \textbf{3.51}  & \textbf{3.60} & \textbf{10.83}\\
    \hline
    \end{tabular}%
  \label{tab:gop}%
\end{table*}%

\begin{table*}[!t]
\scriptsize
\hspace{-1em}
\centering
\caption{\footnotesize{$\Delta \text{PSNR}$ and $\Delta \text{EW-PSNR}$ (dB) of our SGCC approach for HEVC bitstreams with GOP size as 4.}}
\begin{tabular}{|c|c|c|c|c|c|c|c|}
\hline
\multirow{2}[2]{*}{ {Class}} & \multirow{2}[2]{*}{ {Sequence}} & \multicolumn{2}{c|}{QP=32, $\Delta\!C_T\!\!\!=\!\!8\%$$/$QP=22, $\Delta\!C_T\!\!\!=\!\!5\%$} & \multicolumn{2}{c|}{{QP=32, $\Delta\!C_T\!\!\!=\!\!18\%$$/$QP=22, $\Delta\!C_T\!\!\!=\!\!15\%$}} & \multicolumn{2}{c|}{QP=32, $\Delta\!C_T\!\!\!=\!\!23\%$$/$QP=22, $\Delta\!C_T\!\!\!=\!\!20\%$} \\
\cline{3-8}      &             & $\Delta\text{PSNR}$ & $\Delta\text{EW-PSNR}$ & $\Delta\text{PSNR}$ & $\Delta\text{EW-PSNR}$ & $\Delta\text{PSNR}$ & $\Delta\text{EW-PSNR}$ \\
\hline
\multirow{2}[4]{*}{A} & {\textit{Traffic}}    & 0.1117 / 0.0955 & 0.0894 / 0.0594 & 0.4280 / 0.9878 & 0.3564 / 0.5921 & 1.3400 / 6.8979 & 0.8287 / 6.1152\\
\cline{2-8}      & {\textit{PeopleOnStreet}}  & 0.2307 / 0.2076 & 0.0969 / 0.0966 & 0.6619 / 1.3651 & 0.5962 / 0.5960 & 1.3069 / 8.4901 & 0.9758 / 7.0114   \\
\hline
\multirow{3}[6]{*}{B} & {\textit{ParkScene}}  & 0.0934 / 0.0951 & 0.1250  / 0.1256 & 0.5035  / 1.6180 & 0.5662  / 0.9668 & 1.3820  / 7.2926 & 1.0978  / 6.5683   \\
\cline{2-8}      & {\textit{BQTerrace}} & 0.0342 / 0.0320 & 0.0182  / 0.0210 & 0.4270  / 1.7091 & 0.1038  / 0.3968 & 2.1345  / 8.6322 & 1.0791  / 8.9453  \\
\cline{2-8}      & {\textit{Kimono}}   & 0.2852 / 0.1479 & 0.1538  / 0.0870 & 0.6056  / 1.1017 & 0.6826  / 1.1726 & 1.0407  / 5.7960 & 0.9926  / 3.9692   \\
\hline
\multirow{2}[4]{*}{C} & {\textit{RaceHorses}}   & 0.1833 / 0.2869 & 0.1759 / 0.3313 & 0.6622 / 3.2169 & 0.5409 / 2.5636 & 1.5526 / 9.5566 & 1.3170 / 9.4775 \\
\cline{2-8}      & {\textit{PartyScene}}   & 0.0521 / 0.0947 & 0.0457 / 0.2057 & 1.2956 / 5.3079 & 0.4321 / 1.6680 & 3.5283 / 12.3356 & 1.5974 / 9.9862  \\
\hline
\multirow{4}[6]{*}{D} & {\textit{RaceHorses}}   & 0.1585 / 0.1528 & 0.1098 / 0.1814 & 0.4697 / 3.7458 & 0.3922 / 1.8411 & 1.3461 / 9.9917 & 1.1426 / 9.8653  \\
\cline{2-8}      & {\textit{BQSquare}}  & 0.0495 / 0.0701 & 0.0340 / 0.0655 & 1.1450 / 6.5028 & 0.4106 / 4.8822 & 3.2117 / 12.1801 & 2.2608 / 12.2362 \\
\cline{2-8}      & {\textit{BlowingBubbles}}   & 0.0099 / 0.0366 & 0.0128 / 0.0513 & 1.3174 / 8.0826 & 0.8062 / 6.2948 & 4.7982 / 14.8854 & 4.0667 / 14.8432  \\
\cline{2-8}      & {\textit{BasketballPass}}   & 0.1717 / 0.1612 & 0.1171 / 0.1468 & 0.5093 / 1.8775 & 0.4551 / 0.9101 & 1.0825 / 8.2459 & 0.8343 / 8.2901  \\
\hline
\multicolumn{2}{|c|}{{ {Average}}}  &  0.1255  / 0.1255 & 0.0890  / 0.1247 & 0.7296  / 3.2287 & 0.4857  / 1.9895 & 2.0658  / 9.4822 & 1.4721  / 8.8462\\
\hline
\end{tabular}\label{tab:goppsnr}
\end{table*}

\section{Conclusion}
This paper has proposed a decoding complexity control approach (namely SGCC) for HEVC, aiming to reduce HEVC decoding complexity to a target with minimal loss on perceptual quality. We found two ways to reduce the decoding complexity of some CTUs: (1) disabling DF and (2) simplifying MC.  However, disabling DF or simplifying MC may cause some visual quality loss in decoded videos. Thus, the SGCC formulation was proposed to reduce  HEVC decoding complexity to the target, meanwhile minimizing perceptual quality loss. In this paper,  perceptual quality loss was evaluated on the basis of video saliency. For our formulation, the least square fitting on training data was applied to model the relationship between complexity reduction/quality loss and DF disabling/MC simplification. Finally, a potential solution to the proposed formulation was developed, such that SGCC can be accomplished for HEVC decoding. As verified in experimental results, our SGCC approach is efficient in complexity control for HEVC decoding, evaluated in control performance, complexity-distortion performance, fluctuation of quality loss, and subjective quality.

Our work in current form is implemented on HEVC RA bitstreams with hierarchical and open GOP structure. It is an interesting future work to apply our work on other settings, like close GOP structure or LD scenario. Moreover, in current stage, our SGCC approach only concentrates on LCU level complexity control for HEVC decoding. It is another promising future work to control HEVC decoding complexity at
frame level, which may make our SGCC approach more flexible for controlling decoding complexity.

\begin{appendices}
\section{Proof for Lemma 3}
\setcounter{equation}{31}

First, upon (23), $I=\sum_{n=1}^{N}f_n$ holds. Defining $M=\sum_{n=1}^{N}f^{'}_n$, we can turn (24) to
\begin{equation} \label{case2}
 a\cdot\left(\sum_{n=1}^{N}  w_n\cdot f_n \right)+ I\cdot b = a\cdot\left(\sum_{n=1}^{N}  w_n\cdot f^{'}_n \right)+ M\cdot b.
\end{equation}
If $I\geq M$, then $I\cdot b\geq M\cdot b$ exists due to $b>0$. Given $I\cdot b\geq M\cdot b$ and \eqref{case2}, we can obtain (25) because $a>0$. Therefore, for the proof of (25), we only need to prove $I\geq M$.

Next, we prove $I\geq M$ by contradiction as follows.
In the case of $I<M$, we have $\sum_{n=1}^{N} f_n < \sum_{n=1}^{N} f^{'}_n$. Because $f_n=1$ holds if and only if $w_n$ ($\in \left[0,1\right]$) belongs to the smallest $I$ values in $\{w_n\}_{n=1}^{N}$, the following inequality exists,
\begin{equation} \label{r}
\sum_{n=1}^{N} w_n\cdot f_n < \sum_{n=1}^{N} w_n\cdot f^{'}_n.
\end{equation}
When $I<M$ and $b>0$, it is obvious that $I\cdot b<M\cdot b$ holds. Then, given $a>0$, we can obtain
\begin{equation} \label{contra}
a\cdot\left(\sum_{n=1}^{N} w_n\cdot f_n \right)+ I\cdot b < a\cdot\left(\sum_{n=1}^{N} w_n\cdot f^{'}_n \right)+ M\cdot b.
\end{equation}
However, \eqref{contra} contradicts with (\ref{case2}). Hence, the assumption of $I<M$ dose not hold, such that $I\geq M$ can be proved. Finally, the inequality of (25) is achieved.

This completes the proof of Lemma 3. \hfill{$\blacksquare$}

\section{Proof for Lemma 4}
First, $\sum_{n=1}^{N}g_n=N_1+2N_2+3N_3$ holds, since $N_0$, $N_1$, $N_2$ and $N_3$ are the numbers of CTUs at $g_n=0,1,2,3$. As such, the constraint terms of (22-b) and (28) are equivalent. Next, constraint term $\frac{1}{N}\cdot c \cdot (N_1+2N_2+3N_3)$ of (22-b) is independent of $w_n$. Hence, larger $g_n$ should correspond to smaller $w_n$ to make $\sum_{n=1}^N w_n\cdot(h_1\cdot g_n^3+h_2\cdot g_n^2+h_3\cdot g_n)$ minimal. This way, for each combination of $N_0$, $N_1$, $N_2$ and $N_3$, the optimal solution satisfies $\forall w_n \leq w_{n^{'}}, g_n \geq g_{n^{'}}$. Then, the values of $N_0$, $N_1$, $N_2$ and $N_3$ are the variables to be solved for the minimization problem of (22-b). Defining $\{ \widetilde w_n \} _{n=1}^N$ as the ascending sort of $\{w_n \} _{n=1}^N$, the optimal solution $\{ g_n\}_{n=1}^{N}$ towards (22-b) can be written as
\begin{equation} \label{gnn}
 g_n=\left\{
\begin{aligned}
3&,& \quad w_n \leq \widetilde{w}_{N_3} \\
2&,& \quad \widetilde{w}_{N_3+1}\leq w_n \leq \widetilde{w}_{N_3+N_2} \\
1&,& \quad \widetilde{w}_{N_3+N_2+1}\leq w_n \leq \widetilde{w}_{N_3+N_2+N_1} \\
0&,& \quad \widetilde{w}_{N_3+N_2+N_1+1}\leq w_n \leq \widetilde{w}_{N}. \\
\end{aligned}
\right.
\end{equation}
Upon (22-b) and \eqref{gnn}, we can obtain (28). Note that $h_1\cdot g_n^3+h_2\cdot g_n^2+h_3\cdot g_n= 1$ when $g_n=3$, according to (16).

Finally, Lemma 4 can be proved. \hfill{$\blacksquare$}

\section{Proof for Proposition 5}

\begin{figure*}[!t]
\centering
  \subfigure{\includegraphics[width=.5\linewidth]{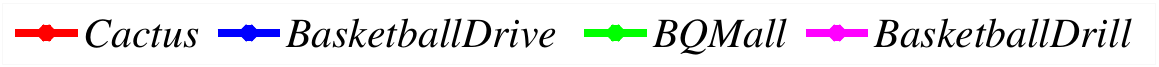}}\\
  \subfigure{\includegraphics[width=.24\linewidth]{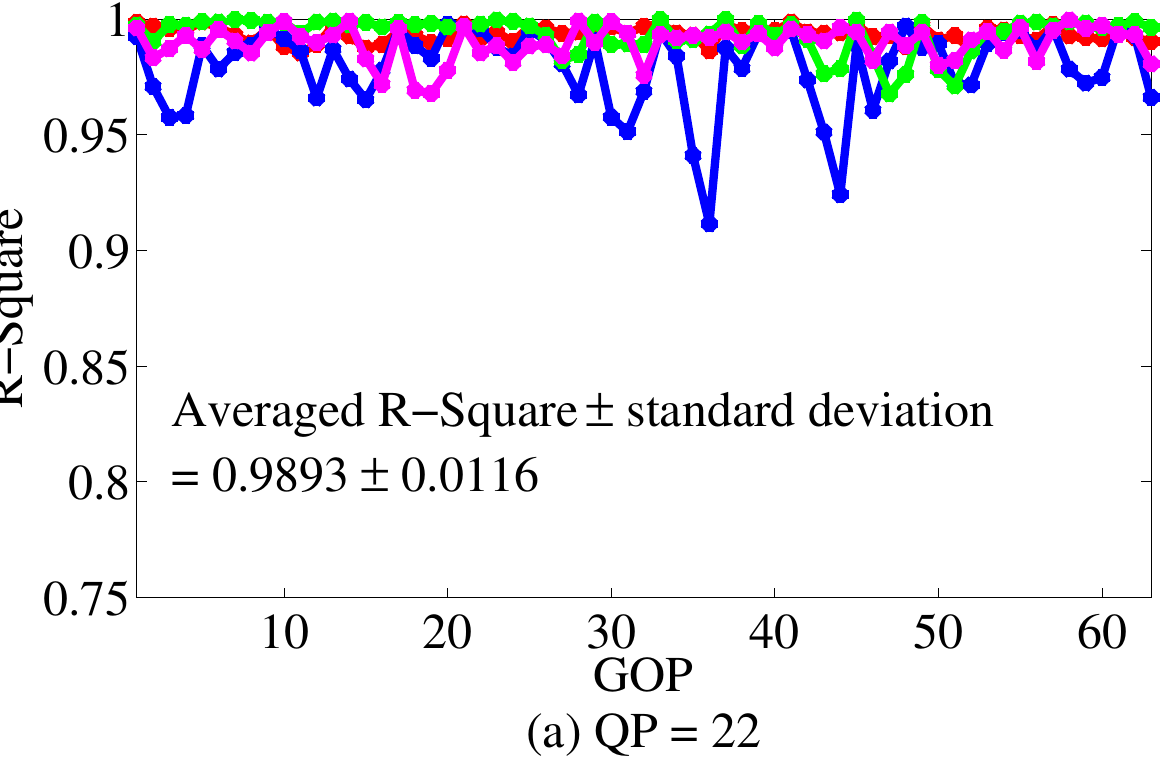}}
  \subfigure{\includegraphics[width=.24\linewidth]{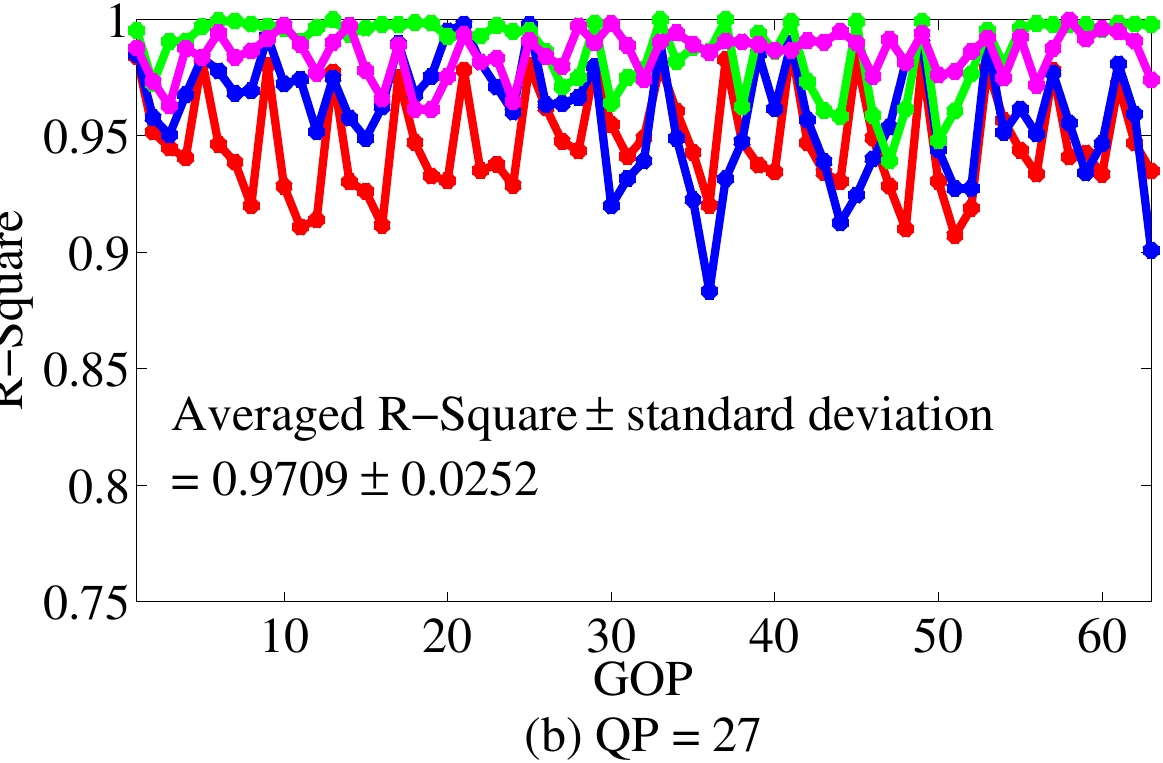}}
  \subfigure{\includegraphics[width=.24\linewidth]{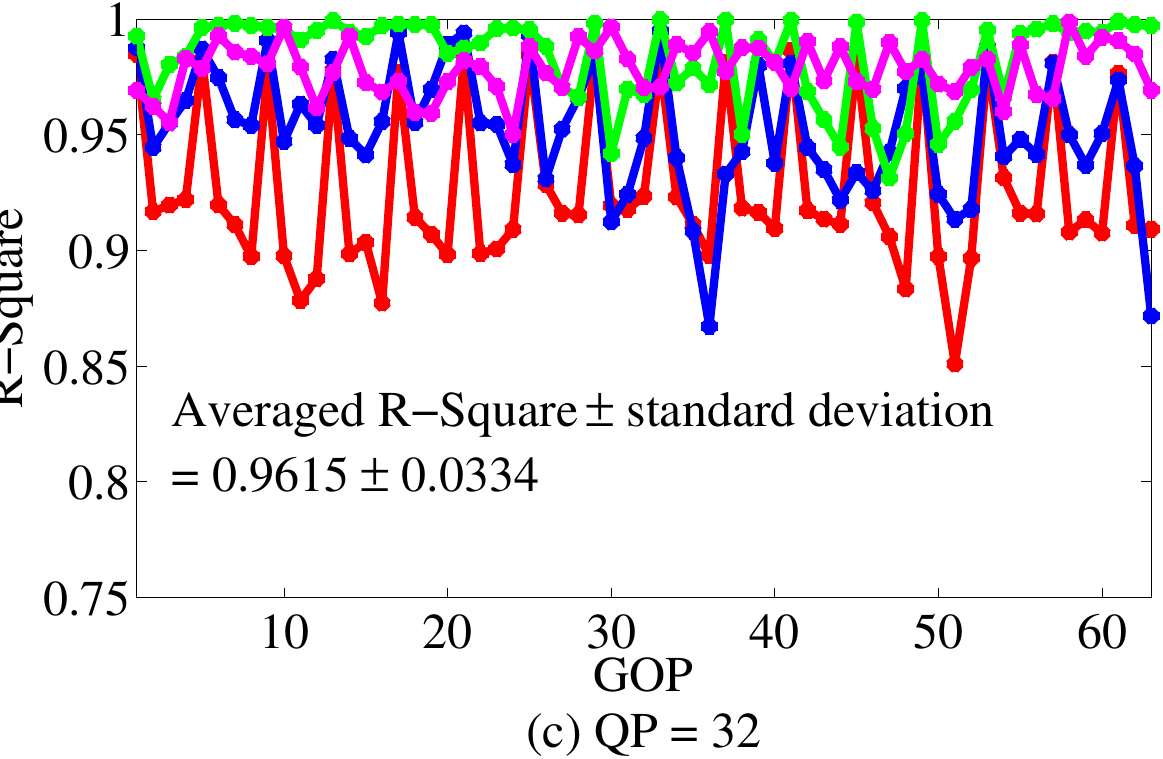}}
  \subfigure{\includegraphics[width=.24\linewidth]{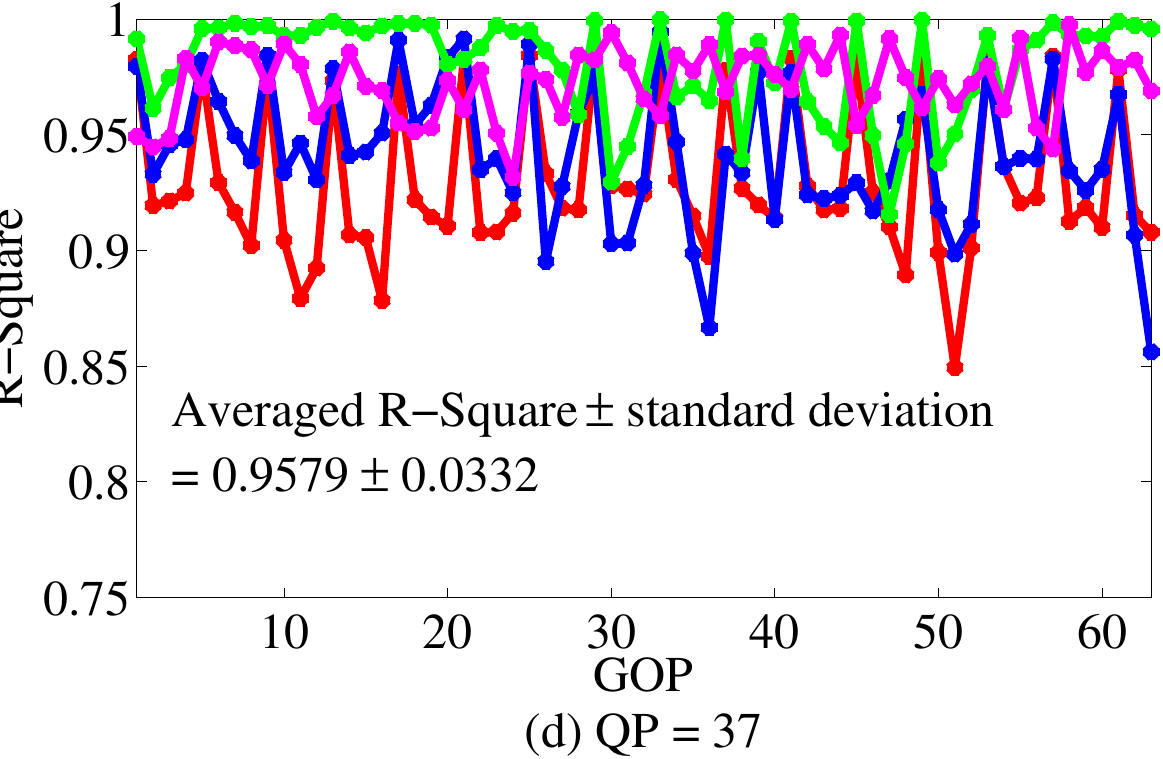}}
  \vspace{-1em}
  \caption{The R-square values for the second-order polynomial regression of $\sum_{n=1}^{N_t}\widetilde w_n = k\cdot N_t^2$. Note that the R-square values of the 3rd frames in each GOP are shown.}\label{fig:sig_w}
\end{figure*}
First, we apply our method of Section II to estimate the saliency values $\{w_n\}_{n=1}^N$ of all CTUs in the four training sequences (the same as Section \uppercase\expandafter{\romannumeral4}) at four QPs (i.e., 22, 27, 32 and 37). Then, at each QP, the values of $\sum_{n=1}^{N_t}\widetilde w_n$ for all possible $N_t \in \{n\}_{n=1}^{N}$ are calculated and recorded at each frame of the four sequences. Recall that $\{\widetilde w_n\}_{n=1}^N$ is the ascending sorted set of saliency values $\{w_n\}_{n=1}^N$ for each frame. Next, we apply the second-order polynomial regression to each frame for modelling the relationship between $N_t$ and $\sum_{n=1}^{N_t}\widetilde w_n$ in form of $\sum_{n=1}^{N_t}\widetilde w_n = k\cdot N_t^2$, where $k$ is the second-order parameter of the regression for a video frame. Note that $k$ is a constant within a video frame, despite being different across frames. The R-square values of such regression can be obtained across different frames of four training sequences. Fig. \ref{fig:sig_w} shows R-square of $\sum_{n=1}^{N_t}\widetilde w_n = k\cdot N_t^2$ regression along with various frames. It is evident that the R-square values are above 0.85 for all frames at four QPs. In addition, the averaged R-square values and their standard deviations are $0.9893\pm 0.0116$, $0.9709\pm 0.0252$, $0.9615\pm 0.0334$ and $0.9579\pm 0.0332$ at QP = 22, 27, 32 and 37. Thus, $\sum_{n=1}^{N_t}\widetilde w_n$ can be well approximated by  $k\cdot N_t^2$.

Finally, Proposition 5 can be proved. \hfill{$\blacksquare$}
\end{appendices}

\ifCLASSOPTIONcaptionsoff
  \newpage
\fi
%
%\begin{thebibliography}{1}
%H.~Kopka and P.~W. Daly, \emph{A Guide to {\LaTeX}}, 3rd~ed.\hskip 1em plus
%\end{thebibliography}
%\bibliographystyle{IEEEbib}
%\vspace{-.5em}
\bibliographystyle{IEEEtran}
\bibliography{references}

\begin{IEEEbiography}[{\includegraphics[width=1\linewidth]{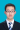}}]{Ren Yang}

received the B.S. degree from Beihang University in 2012. He is currently pursuing the M.S. degree at the School of Electronic and Information Engineering, Beihang University, Beijing, China. His research interests mainly include computer vision and video coding.

\end{IEEEbiography}

\begin{IEEEbiography}[{\includegraphics[width=1\linewidth]{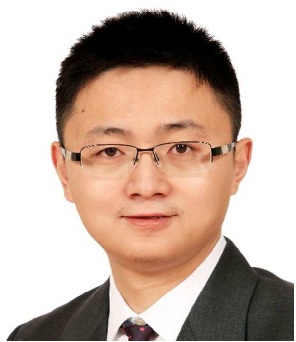}}]{Mai Xu}

(M'10, SM'16) received B.S. degree from Beihang University in 2003, M.S. degree from Tsinghua University in 2006 and Ph.D degree from Imperial College London in 2010. From 2010-2012, he was working as a research fellow at Electrical Engineering Department, Tsinghua University. Since Jan. 2013, he has been with Beihang University as an Associate Professor. During 2014 to 2015, he was a visiting researcher of MSRA. His research interests mainly include image processing and computer vision.  He has published more than 60 technical papers in international journals and conference proceedings, e.g., IEEE TIP, CVPR and ICCV. He is the recipient of best paper awards of two IEEE conferences.

\end{IEEEbiography}

\begin{IEEEbiography}[{\includegraphics[width=1\linewidth]{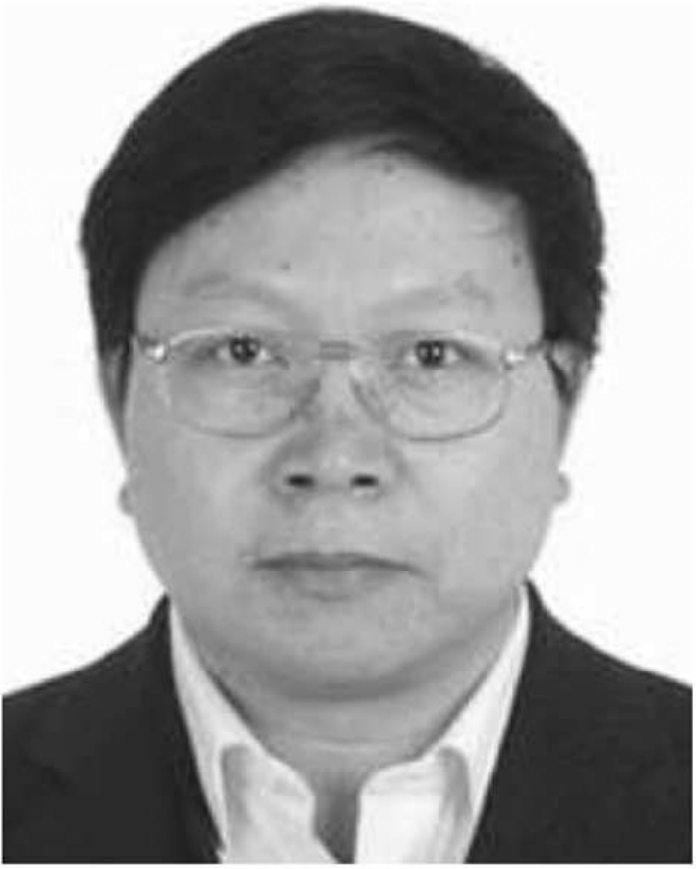}}]{Zulin Wang}

(M'14) received the B.S. and M.S. degrees in electronic engineering from Beihang University, in 1986 and 1989, respectively. He received his Ph.D. degree at the same university in 2000. His research interests include image processing, electromagnetic countermeasure, and satellite communication technology. He is author or co-author of over 100 papers and holds 6 patents, as well as published 2 books in these fields. He has undertaken approximately 30 projects related to image/video coding, image processing, etc.
\end{IEEEbiography}

\begin{IEEEbiography}[{\includegraphics[width=1\linewidth]{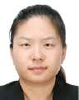}}]{Yiping Duan}

received the B. S. degree at the school of computer science and technology, Henan Normal University, Xinxiang, China, in 2010, and Ph. D. from the school of computer science and technology, Xidian University, Xi¡¯an, China, in 2016.She is currently working towards postdoctoral research with the Department of Electronic Engineering, Tsinghua University. Her current research interests include semantic mining, machine learning and remote sensing image processing.

\end{IEEEbiography}

\begin{IEEEbiography}[{\includegraphics[width=1\linewidth]{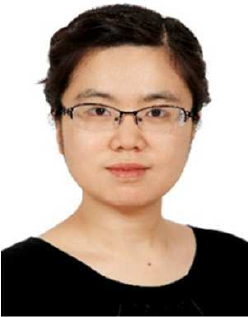}}]{Xiaoming Tao}
(M'09) received the B.E. degree from the School of Telecommunications Engineering, Xidian University in 2003, and Ph. D. from the Department of Electronic Engineering, Tsinghua University, Beijing, China, in 2008. She is currently an Associate Professor with the Department of Electronic Engineering, Tsinghua
University. Her research interests include wireless communications and networking, and multimedia
signal processing.
\end{IEEEbiography}

\end{document}